\documentclass[fleqn,usenatbib]{mnras}

\pdfoutput=1

\usepackage{newtxtext,newtxmath}

\usepackage[T1]{fontenc}
\usepackage{ae,aecompl}

\usepackage{graphicx}	
\usepackage{amsmath}	
\newcommand{\eg}[0]{$\textnormal{e.g. }$}
\newcommand{\ie}[0]{$\textnormal{i.e. }$}
\newcommand{\Msun}[0]{\,\textnormal{M}_{\textnormal{$\odot$}}}

\newcommand{\Zsun}[0]{\,\textnormal{Z}_{\textnormal{$\odot$}}}
\newcommand{\Zg}[0]{Z_{\textnormal{g}}}
\newcommand{\Zgt}[0]{$Z_{\textnormal{g}}$}
\newcommand{\Zs}[0]{Z_{*}}
\newcommand{\Zst}[0]{$Z_{*}$}
\newcommand{\Zh}[0]{Z_{\textnormal{h}}}
\newcommand{\Zht}[0]{$Z_{\textnormal{h}}$}

\newcommand{\ZICMt}[0]{$Z_{\textnormal{\textsc{ICM}}}$}
\newcommand{\Te}[0]{$T\sub{e}$}

\newcommand{\HII}[0]{H\textsc{ii}}
\newcommand{\tn}[1]{\textnormal{#1}}
\newcommand{\sub}[1]{_{\textnormal{#1}}}

\newcommand{\error}[1]{\tn{\scriptsize{$\pm #1$}}}
\newcommand{\lgal}[0]{\textsc{L-Galaxies}}
\newcommand{\lgaltt}[0]{\textsc{L-Galaxies 2020}}
\newcommand{\logM}[0]{log$_{10}(M_{*}/\Msun)$}
\newcommand{\logMm}[0]{\tn{log}_{10}(M_{*}/\Msun)}
\newcommand{\Reff}[0]{$R\sub{e}$}
\newcommand{\Reffm}[0]{R\sub{e}}
\newcommand{\logten}[0]{\tn{log}_{10}}

\newcommand{\NeIII}[0]{[Ne\textsc{iii}]}
\newcommand{\NII}[0]{[N\textsc{ii}]}
\newcommand{\OII}[0]{[O\textsc{ii}]}
\newcommand{\OIII}[0]{[O\textsc{iii}]}
\newcommand{\Ha}[0]{H$\alpha$}
\newcommand{\Hb}[0]{H$\beta$}
\newcommand{\SII}[0]{[S\textsc{ii}]}

\title[Evolution of metallicity profiles]{\textsc{L-GALAXIES} 2020: The evolution of radial metallicity profiles and \\global metallicities in disc galaxies}

\author[Robert M. Yates et al.]{Robert M.~Yates,$^{1}$\thanks{E-mail: \href{mailto:robyates@mpa-garching.mpg.de}{robyates@mpa-garching.mpg.de}}
Bruno M. B. Henriques,$^{2}$
Jian Fu,$^{3}$
Guinevere Kauffmann,$^{1}$
Peter A. Thomas,$^{4}$
\newauthor
Qi Guo,$^{5}$
Simon D. M. White$^{1}$
and 
Patricia Schady$^{6}$
\\
$^{1}$Max-Planck-Institut f{\"u}r Astrophysik, Karl-Schwarzschild-Stra\ss{}e 1, 85741, Garching, Germany \\
$^{2}$Institute for Astronomy, ETH Zurich, CH-8093 Zurich, Switzerland\\
$^{3}$Key Laboratory for Research in Galaxies and Cosmology, Shanghai Astronomical Observatory, CAS, 80 Nandan Road, Shanghai, 200030, China\\
$^{4}$Astronomy Centre, University of Sussex, Falmer, Brighton BN1 9QH, UK\\
$^{5}$Partner Group of the Max-Planck-Institut f{\"u}r Astrophysik, National Astronomical Observatories, Chinese Academy of Sciences, Beijing, 100012, China\\
$^{6}$Department of Physics, University of Bath, Claverton Down, Bath, BA2 7AY, UK
}

\date{Accepted XXX. Received YYY; in original form ZZZ}

\pubyear{2021}

\begin{document}
\label{firstpage}
\pagerange{\pageref{firstpage}--\pageref{lastpage}}
\maketitle

\begin{abstract}
We present a modified version of the \lgaltt{} semi-analytic model of galaxy evolution, which includes significantly increased direct metal enrichment of the circumgalactic medium (CGM) by supernovae (SNe). These more metal-rich outflows do not require increased mass-loading factors, in contrast to some other galaxy evolution models. This modified \lgaltt{} model is able to simultaneously reproduce the gas-phase metallicity (\Zgt{}) and stellar metallicity (\Zst{}) radial profiles observed in nearby disc galaxies by MaNGA and MUSE, as well as the observed mass -- metallicity relations for gas and stars at $z=0$ and their evolution back to $z\sim{}2-3$. A direct CGM enrichment fraction of $\sim{}90$ per cent for SNe-II is preferred. We find that massive disc galaxies have slightly flatter \Zgt{} profiles than their lower-mass counterparts in \lgaltt{}, due to more efficient enrichment of their outskirts via inside-out growth and metal-rich accretion. Such a weak, positive correlation between stellar mass and \Zgt{} profile slope is also seen in our MaNGA-DR15 sample of 571 star-forming disc galaxies, although below $\logMm{}\sim{}10.0$ this observational result is strongly dependent on the metallicity diagnostic and morphological selection chosen. In addition, a lowered maximum SN-II progenitor mass of $25\Msun$, reflecting recent theoretical and observational estimates, can also provide a good match to observed \Zgt{} and \Zst{} profiles at $z=0$ in \lgaltt{}. However, this model version fails to reproduce an evolution in \Zgt{} at fixed mass over cosmic time, or the magnesium abundances observed in the intracluster medium (ICM).
\end{abstract}

\begin{keywords}
methods: analytical -- methods: data analysis -- galaxies: abundances -- galaxies: evolution
\end{keywords}



\color{black}

\section{Introduction}\label{sec:Intro}
The evolution of the metal content in and around galaxies remains a key area of contention in our understanding of galaxy evolution. On the observational side, discrepancies and uncertainties in metallicity measurements at both low and high redshift have led to conflicting interpretations for the interstellar medium (ISM) (\eg{}\citealt{Kewley&Dopita02,Yates+12,Sanchez+14,Belfiore+17,Erroz-Ferrer+19}), stellar populations (\eg{}\citealt{Maraston05,Lonoce+20}), and the circumgalactic medium (CGM) (\eg{}\citealt{Tumlinson+17,Werner&Mernier20,Peroux&Howk20}). On the modelling side, degeneracies in the physical prescriptions implemented, and the tendency to focus on only a single galactic component or redshift, has hampered our ability to draw robust, definitive conclusions (\eg{}\citealt{Fu+12,Somerville+15,Somerville&Dave15,Collacchioni+18,Mitchell+20a}). Therefore, in order to make further progress in the field of galactic chemical evolution (GCE), we must look to combine accurate observational data from a range of phases and spatial scales with detailed theoretical models that allow the simultaneous modelling of diverse galaxy populations.

To this end, we present a study of the \lgaltt{} semi-analytic model, which allows the self-consistent modelling of galaxies and their environments back to high redshift, including the internal evolution of their gas and stellar discs. This model is compared to a range of metallicity observations of the ISM, stars, CGM, and ICM, in order to more accurately constrain the relative importance of the various physical processes driving metallicity evolution in and around galaxies.

This paper is organised as follows: In Section \ref{sec:L-Galaxies}, we introduce our modified \lgaltt{} model, and compare it to the default version presented in \citet{Henriques+20}. In Section \ref{sec:MaNGA}, we present the MaNGA sample used as our main low-redshift observational data set. In Section \ref{sec:Results}, we discuss the comparison between \lgaltt{} and various observations at low and high redshift, both for global metallicities and metallicity profiles. In Section \ref{sec:Comparison to other models}, we contrast the findings from \lgaltt{} with those from other galaxy evolution models from the literature. In Section \ref{sec:Conclusions}, we summarize our conclusions.

\section{\textsc{L-Galaxies 2020}}\label{sec:L-Galaxies}
\lgaltt{} is a semi-analytic model of galaxy evolution, built to run on the dark matter (DM) subhalo merger trees of N-body simulations of cosmic structure formation. In this work, we apply \lgaltt{} to the merger trees from the $(480.3\ \tn{Mpc}/h)^{3}$ \textsc{Millennium} and $(96.1\ \tn{Mpc}/h)^{3}$ \textsc{Millennium-II} simulations \citep{Springel+05,Boylan-Kolchin+09}. This enables the study of millions of model galaxies in the mass range $7.0 \lesssim \logMm{} \lesssim 12.0$, evolved from $z=56$ to the present day in a \textit{Planck-I} cosmology (see \citealt{Angulo&White10,Angulo&Hilbert15}).

This version of the \lgal{} model is the latest in a long line of major releases (\eg{}\citealt{Springel+01,Springel+05,DeLucia&Blaizot07,Guo+11,Henriques+15}). The new model is discussed in detail in \citet{Henriques+20} and in the supplementary material available online.\footnote{Supplementary material and model output catalogues from \lgaltt{} are available at https://lgalaxiespublicrelease.github.io/} Below, we give a brief overview of the key aspects most relevant to this work.

In addition to the existing implementations of gas cooling, SN feedback, AGN feedback, and other key processes, \lgaltt{} improves on the previous version of the model \citep{Henriques+15} by including molecular hydrogen (H$_{2}$) formation \citep{Fu+10}, detailed chemical enrichment \citep{Yates+13}, and radially-resolved gas and stellar discs \citep{Fu+13}. This opens-up a whole new dimension of study into galaxy evolution, by allowing comparison to, and interpretation of, the latest observations of gas and stellar properties within galaxies from integral field units (IFUs).


In \lgaltt{}, gas and stellar discs are divided into 12 concentric annuli (or `rings') of fixed radius and width. As explained by \citeauthor{Henriques+20} (2020, section 2.1), the outer edge of the $i$th ring is given by $r\sub{i} = 0.01\cdot{}2^{i}\,h^{-1}$ kpc, such that higher spatial resolution is obtained at lower radii. For our chosen cosmology, $r\sub{i}$ ranges from 60.24 kpc for the outermost ring to 0.03 kpc for the innermost ring, meaning that \lgaltt{} resolves discs down to sub-kpc resolution in the centres of galaxies. The gas disc represents the cold interstellar medium (ISM) of galaxies, consisting of \HII{} regions, atomic, and molecular gas. Material is allowed to flow between rings in the gas disc following a linear scaling between inflow velocity and radius, $v\sub{inflow}=\alpha\sub{inflow}\,r$, where $\alpha\sub{inflow}$ is the gas inflow parameter. Gas can also be expelled from discs via SN feedback occurring in each ring (see Section \ref{sec:GCE parameters}), and can cool onto the galaxy from the hot circumgalactic medium (CGM) at all radii. This accretion onto the disc is assumed to follow an exponential radial profile (see \citealt{Henriques+20}, section 2.2). The CGM in \lgaltt{}, which encompasses all the hot gas surrounding a galaxy out to its virial radius (but not the material fully ejected out of the halo by feedback), is not spatially resolved. Consequently, the gas accreted onto discs is assumed to have a uniform metallicity. Likewise, the material within each ring in discs is assumed to be fully mixed, in qualitative agreement with observations of homogeneous azimuthal metal distributions within the ISM (\eg{}\citealt{Li+13,Kreckel+16,Kreckel+20}).


The galactic chemical enrichment (GCE) scheme implemented into \lgaltt{} is explained by \citeauthor{Henriques+20} (2020, section 2.4). In brief, the ages of AGB stars, SN-Ia progenitors, and SN-II progenitors from each stellar population formed are tracked, and their mass- and metallicity-dependent ejecta released into the ISM and CGM at the end of their lifetimes. The apportionment of this metal-rich material between the ISM and CGM is set by the model GCE parameters, which are discussed in the following section. The yield tables used in \lgaltt{} are taken from \citet{Marigo01} for AGB stars, \citet{Thielemann+03} for SNe-Ia, and \citet{Portinari+98} for SNe-II. A power-law delay-time distribution (DTD) with a slope of $-1.12$ is assumed for SNe-Ia, following \citet{Maoz+12}. The mass returned by SNe is used to calculate the amount of energy they deposit. This allows us to model a form of extended SN feedback in \lgaltt{}, whereby stars from the same stellar population contribute to the reheating and ejection of gas at different times, depending on their mass- and metallicity-dependent lifetimes. This SN feedback scheme is discussed further in Sections \ref{sec:reheating and ejection rates} and \ref{sec:mass-loading factors} below.

H$_{2}$ formation in \lgaltt{} follows the gas partitioning scheme developed by \citet{Krumholz+09,McKee&Krumholz10}. In this formalism, the H$_{2}$ mass fraction ($\mu\sub{H2}$) in the ISM depends on the local gas density and metallicity (see \citealt{Henriques+20}, section 2.2.3). Effectively, $\mu\sub{H2}$ transitions from 0.0 at $\Sigma\sub{ISM} \lesssim 30\,\Msun\tn{pc}^{-2}$ to 1.0 at $\Sigma\sub{ISM} \gtrsim 100\,\Msun\tn{pc}^{-2}$, with only a weak secondary dependence on metallicity. The star-formation rate density in each ring in \lgaltt{} is then calculated from the local H$_{2}$ density and the DM subhalo dynamical time, $\Sigma\sub{\textsc{SFR}} = \alpha\sub{\textsc{SFR}}\,\Sigma\sub{H2}/t\sub{dyn}$, where $\alpha\sub{\textsc{SFR}} = 0.06$ is the assumed dimensionless star formation efficiency (see \citealt{Henriques+20}, section 2.2.4). The $\alpha\sub{\textsc{SFR}}$ parameter, along with a number of other key physical parameters, is self-consistently constrained in \lgaltt{} using the Monte Carlo Markov Chain (MCMC) formalism developed by \citet{Henriques+09,Henriques+15}.



\subsection{Modifications to the GCE parameters}\label{sec:GCE parameters}
The parameters controlling chemical enrichment in \lgaltt{} have been modified in this work to improve the correspondence with metallicity observations on global and sub-galactic scales.

The chief modification we make is to significantly increase the amount of material released by SNe which is allowed to directly enrich the CGM, without first mixing with the ambient ISM. This, along with the amount of direct CGM enrichment by AGB stars, is parameterised in the model by the $f\sub{SNII,hot}$, $f\sub{SNIa,hot}$, and $f\sub{AGB,hot}$ parameters (referred to collectively hereafter as the $f\sub{hot}$ parameters).

Table \ref{tab:GCE parameters} presents the values of the GCE parameters in both the default \lgaltt{} model presented in \citet{Henriques+20} (hereafter, the `default model') and the `modified model' introduced here. Both $f\sub{SNII,hot}$ and $f\sub{SNIa,hot}$ have been increased in the modified model to 90 and 80 per cent, respectively. This has been done to better match the normalisation of the gas-phase metallicity (\Zgt{}) radial profiles seen in nearby galaxies (see Section \ref{sec:Profiles at z=0}), mimicking the metal-rich galactic outflows driven by SNe that are seen in star-forming galaxies both in observations (\eg{}\citealt{Martin+02,Strickland+04,Tumlinson+11}) and hydrodynamical models (\eg{}\citealt{Gibson+13,Li+17,Emerick+20a}). Typically, SNe-Ia are assumed to pollute the hot CGM more efficiently than SNe-II. However, given that a significant fraction of SNe-Ia are also expected to explode promptly (see \eg{}\citealt{Yates+13,Maoz+14}), and that much of the ejecta from SNe-II is expected to deposit into hot, low-density bubbles rather than cold, dense ISM, before outflowing (see \eg{}\citealt{Gatto+17}), it is perhaps not so surprising that the values of $f\sub{SNII,hot}$ and $f\sub{SNIa,hot}$ required here are similar to each other. We find that values below $\sim{}75$ per cent return gas and stellar metallicities in galaxies that are too high at $z=0$, ISM metal enrichment rates that are too high at early times, and light $\alpha$ element abundances in the ICM that are too low by $z=0$, compared to the observations considered in this work.

The $f\sub{AGB,hot}$ parameter has been increased only modestly to 25 per cent in the modified model, in keeping with the expectation that AGB wind ejecta are predominantly released at low velocity and after SNe from the same generation of stars have exploded. More of the AGB ejecta material therefore remains available for immediate mixing with the cold ISM (see \eg{}\citealt{Emerick+18}).

A significant increase in a unified $f\sub{hot}$ parameter was also recommended by \citet{Fu+13} when studying the \citet{Guo+11} version of \lgal{}, in order to explain flat metallicity gradients in galaxies above $\logMm{}\sim{}10$. That version of the model did not contain the detailed GCE scheme \citep{Yates+13} or improvements to gas reincorporation \citep{Henriques+15} present in \lgaltt{}, and was not compared to higher-redshift observational data or IFU data at low redshift.

\begin{table}
\centering
\begin{tabular}{lcc}
\hline \hline
GCE parameters & Default model & Modified model \\
\hline
$f\sub{SNII,hot}$ & 0.3 & 0.9 \\
$f\sub{SNIa,hot}$ & 0.3 & 0.8 \\
$f\sub{AGB,hot}$ & 0.0 & 0.25 \\
$A\sub{SNIa}$ & 0.04 & 0.035 \\
$\alpha\sub{inflow}$ & 1.0 & 0.6 \\
\hline \hline
\end{tabular}
\caption{The GCE parameter values chosen in \lgaltt{} for the default model presented in \citet{Henriques+20} and the modified model presented here. \textit{Rows 1-3:} The fraction of material ejected by SNe-II, SNe-Ia, and AGB stars which is directly added to the hot CGM. \textit{Row 4:} The fraction of stellar objects between 3 and $16\Msun$ in each stellar population that are assumed to be SN-Ia progenitor systems (see section 4 of \citealt{Yates+13}). \textit{Row 5:} The gas inflow parameter (in km s$^{-1}$ kpc$^{-1}$), which sets the speed of inflow within galaxy discs (see section 2.3.3 of \citealt{Fu+13}).}
\label{tab:GCE parameters}
\end{table}

The final two GCE parameters listed in Table \ref{tab:GCE parameters} have a less significant impact on general galaxy evolution in \lgaltt{}. The $A\sub{SNIa}$ parameter, which represents the fraction of stellar objects of mass $3-16\Msun$ per stellar population which produce SNe-Ia, has been lowered slightly from 0.04 to 0.035. In combination with a ratio of $f\sub{SNIa,hot}/f\sub{SNII,hot} < 1$, this enables \lgaltt{} to maintain a good match to observations of the alpha enhancements in early-type galaxies (ETGs) at low redshift (see \citealt{Yates+13}, section 6.3). This is necessary in response to the increase in the value of the $f\sub{hot}$ parameters in the modified model. The gas inflow parameter, $\alpha\sub{inflow}$, has also been lowered to 0.6 km/s/kpc, in order to slow the flow of gas into the centres of galaxies, reducing the amount of central star formation and hence helping to maintain the same HI mass function and central gas densities as seen in the default model (see Section \ref{sec:General galaxy properties}). This produces inflow speeds of $<5$ km/s in the inner $\sim{}8$ kpc of galaxy discs, in good correspondence with high-resolution simulations of Milky-Way-sized galaxies (Okalidis et al., in prep.).

\subsubsection{Reheating and ejection rates}\label{sec:reheating and ejection rates}
Varying the $f\sub{hot}$ parameters in \lgaltt{} not only changes the metal content in and around galaxies, but also the amount of energy available for gas reheating and ejection. As explained in section 2.4.1 of \citet{Henriques+20} and in more detail in section S1.10 of the supplementary material, the amount of energy available for reheating gas from the ISM to the CGM due to stellar feedback is given by
\begin{equation}\label{eqn:Reheating energy}
    \Delta{}E\sub{reheat} = \tn{min}\,\left[\frac{1}{2}\,\epsilon\sub{disc}\,\Delta{}M\sub{ret,ISM}V^{2}\sub{vir}\ \ ,\ \ \Delta{}E\sub{SN,ISM}\right]\;\;,
\end{equation}
where $\epsilon\sub{disc}$ is the reheating efficiency parameter,\footnote{$\epsilon\sub{disc}$ can exceed 1.0 and acts in Eqn. \ref{eqn:Reheating energy} as a proportionality factor between the mass reheated by stellar feedback and the mass returned to the ISM by stars. It is dependent on the maximum rotation velocity of the subhalo, $V\sub{max}$, such that it's value is typically larger in low-mass galaxies (see section S1.10 in the supplementary material).} $V\sub{vir}$ is the virial velocity of the galaxy's DM subhalo, $\Delta{}M\sub{ret,ISM}$ is the mass returned by SNe and stellar winds to the ISM, given by
\begin{align}\label{eqn:Mass returned to ISM}
\nonumber \Delta{}M\sub{ret,ISM}
\nonumber & = (1-f\sub{SNII,hot})M\sub{ret,SNII}\\
\nonumber & + (1-f\sub{SNIa,hot})M\sub{ret,SNIa}\\
    & + (1-f\sub{AGB,hot})M\sub{ret,AGB}\;\;,
\end{align}
and $\Delta{}E\sub{SN,ISM}$ is the total energy deposited into the ISM by SNe (also proportional to $\Delta{}M\sub{ret,ISM}$). We can therefore see that an increase in the $f\sub{hot}$ parameters leads to a decrease in $\Delta{}E\sub{reheat}$. 

In turn, the total energy available for ejection of gas out of the CGM into an ejecta reservoir is given by
\begin{equation}\label{eqn:Ejection energy}
    \Delta{}E\sub{eject} = \epsilon\sub{halo}\,(\Delta{}E\sub{SN,ISM} + \Delta{}E\sub{SN,CGM}) - \Delta{}E\sub{reheat}\,\;\;,
\end{equation}
where $\epsilon\sub{halo}$ is the ejection efficiency parameter\footnote{$\epsilon\sub{halo}$ is simply the fraction of available SN energy that is allowed to couple with the CGM to drive outflows. It is effectively saturated at 1.0 for all subhaloes in \lgaltt{} (see section S1.10 in the supplementary material).} and $\Delta{}E\sub{SN,CGM}$ is the total energy deposited directly into the CGM by SNe, which depends on
\begin{align}\label{eqn:Mass returned to CGM}
\nonumber \Delta{}M\sub{ret,CGM}
\nonumber & = f\sub{SNII,hot}\,M\sub{ret,SNII}\\
\nonumber & + f\sub{SNIa,hot}\,M\sub{ret,SNIa}\\
    & + f\sub{AGB,hot}\,M\sub{ret,AGB}\;\;.
\end{align}

\begin{figure}
\centering
 \includegraphics[angle=0,width=0.8\linewidth]{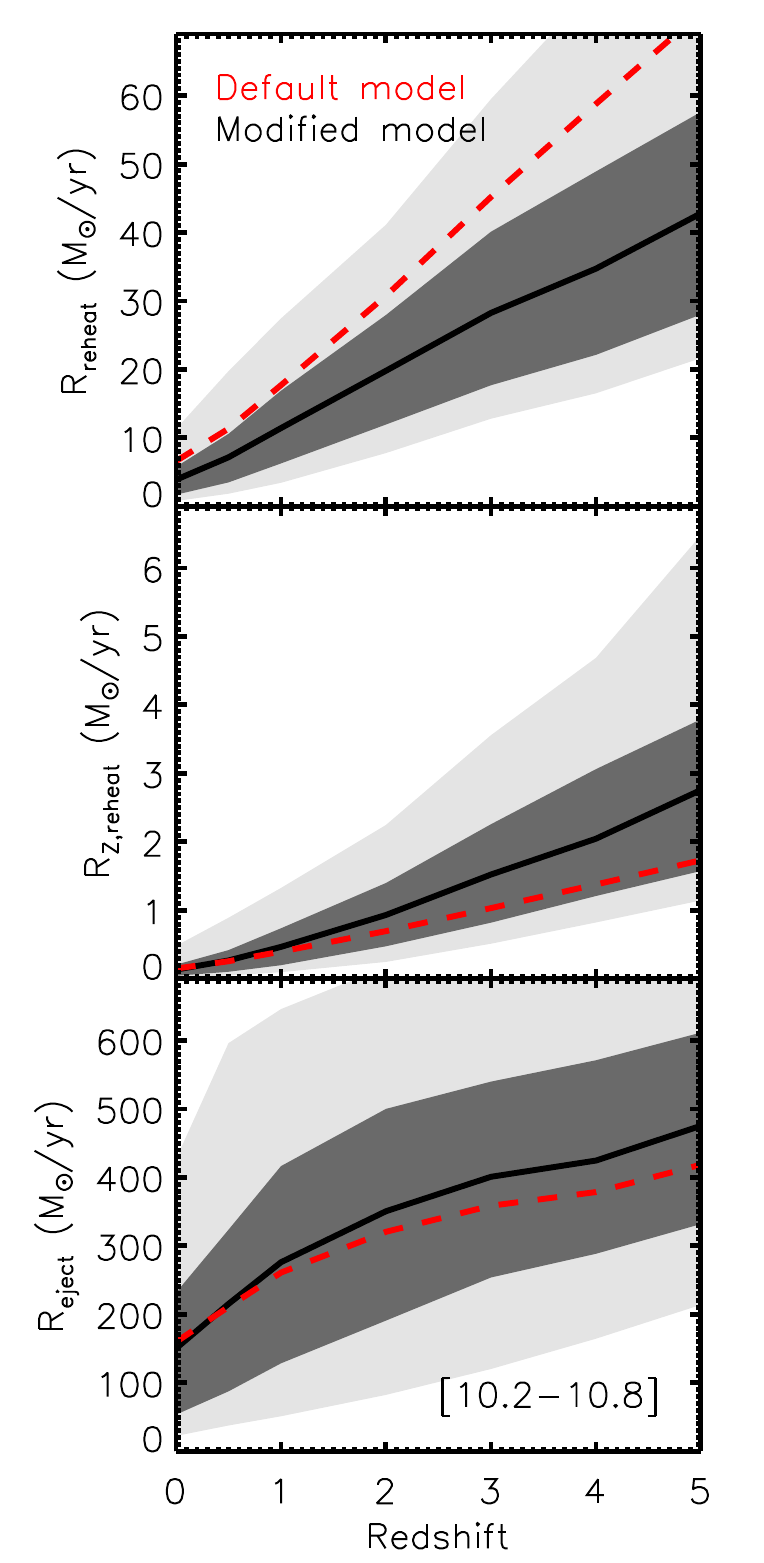}
 \caption{\textit{Top panel:} The mean reheating rate of gas from the ISM into the CGM for star-forming galaxies with $10.2 < \logMm{} < 10.8$ selected at various redshifts. \textit{Middle panel:} The same as the top panel, but for metals only. \textit{Bottom panel:} The mean ejection rate of gas from the CGM into an external ejecta reservoir. In all panels, dashed red lines denote the default model, and solid black lines denote the modified model. Dark and light grey regions represent the 16th-84th and 2nd-98th percentiles ranges for the modified model, respectively.}
 \label{fig:Reheating_and_ejection_rates}
\end{figure}

We can therefore see that the model partitions the total stellar feedback energy available between reheating ISM gas and ejecting CGM gas. Consequently, any reduction in the reheating energy leads to a complementary increase in the ejection energy. Even when the reheating energy is maximal (\ie{}when $\Delta{}E\sub{reheat} = \Delta{}E\sub{SN,ISM}$), $\Delta{}E\sub{eject}$ can still be increased due to the dependence of $\Delta{}E\sub{SN,CGM}$ on the $f\sub{hot}$ parameters.

Fig \ref{fig:Reheating_and_ejection_rates} illustrates this partitioning of SN energy in \lgaltt{} by showing the mean reheating and ejection rates for model star-forming galaxies of Milky-Way mass [$10.2 < \logMm{} < 10.8$], selected at various redshifts. Red dashed lines represent the default model, and solid black lines and shaded regions represent the modified model. For the reheating rates, both components of galactic outflows are considered, namely (a) the SN ejecta itself and (b) the ambient ISM gas entrained in the SN-driven winds. We can see that the increased direct CGM enrichment in the modified model reduces the ISM reheating rate (top panel) by a factor of $\sim{}1.6$. In turn, the rate of ejection of hot gas out of the CGM (bottom panel) increases, because more energy is now available to drive outflows out of the subhalo. 

The middle panel of Fig. \ref{fig:Reheating_and_ejection_rates} shows the reheating rate for metals only, again including both metals released from SNe and those entrained from the ISM. Metals are assumed to be fully mixed in each radial ring of the gas disc, so the entrained component of an outflow has the metallicity of this local ISM. This panel illustrates that more metal-rich galactic winds are present in the modified model, even though the \textit{total} amount of material in these winds is actually lower (see top panel). This feature of the modified \lgaltt{} model distinguishes it from some other recent galaxy evolution models (see Section \ref{sec:Comparison to other models}), and is also reflected in the mass-loading factors discussed in the following section.
\subsubsection{Mass-loading factors}\label{sec:mass-loading factors}
\begin{figure}
\centering
 \includegraphics[angle=0,width=0.8\linewidth]{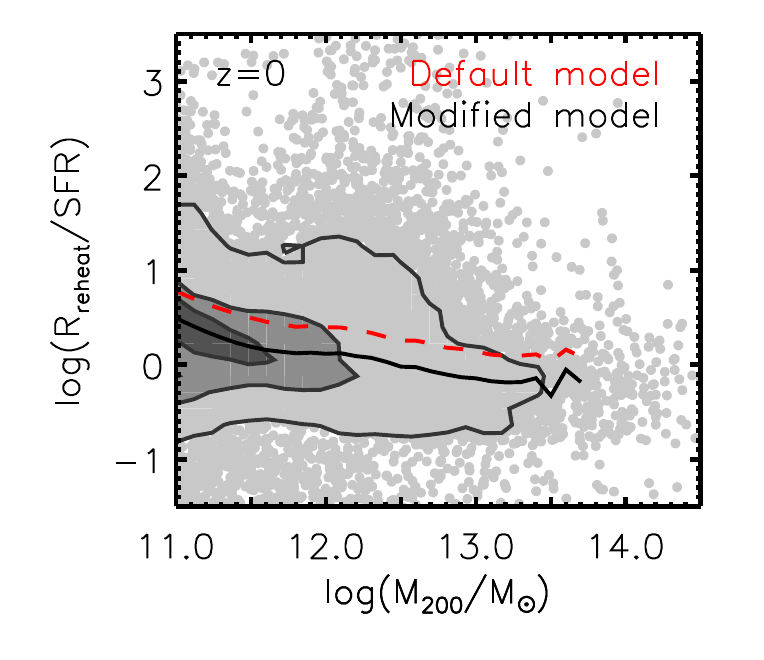}
 \caption{Mass-loading factors for reheated material as a function of $M_{200}$ for all $z=0$ galaxies with non-zero SFRs. Grey contours and points represent galaxies in the modified model. The red dashed and black solid lines represent the median relations from the default and modified models, respectively.}
 \label{fig:mass_loading_factors}
\end{figure}

Although the values of the $f\sub{hot}$ parameters are the same for galaxies of all masses in \lgaltt{}, we note that the mass-loading factors for reheated and ejected material (\ie{}$R\sub{reheat}/$SFR and $R\sub{eject}/$SFR) are still weakly dependent on mass at $z=0$. Fig. \ref{fig:mass_loading_factors} shows $R\sub{reheat}/$SFR as a function of $M_{200}$ for all model galaxies at low redshift. The mass dependence seen is predominantly due to the inverse dependency of the reheating efficiency on $V\sub{max}$. Fig. \ref{fig:mass_loading_factors} also illustrates how the amount of mass loading in the modified model is lower than in the default model, despite the increased metal ejection, as mentioned in the section above.

The reheating mass-loading factors as a function of mass in \lgaltt{} are in good agreement with those found by \citet{Mitchell+20a} for the EAGLE hydrodynamical model. We note that, due to the incorporation of (a) delayed mass return from stars, and (b) non-zero direct enrichment of the CGM, the true mass-loading factors in \lgaltt{} are no longer simply equivalent to the $\epsilon\sub{disc}$ parameter. These mass-loading factors are also in good qualitative agreement with those calculated for observed quiescent galaxies at $z\sim{}0.39-0.54$ by \citet{Leethochawalit+19}.

\subsection{General galaxy properties}\label{sec:General galaxy properties}
The main results from \lgaltt{} concerning the general galaxy population were discussed in detail in \citet{Henriques+20}. Here, we present some of the main relations again, to demonstrate that the changes to the GCE parameters discussed in Section \ref{sec:GCE parameters} have not significantly altered the model's match to its key observational constraints.

\begin{figure*}
\centering
\begin{tabular}{@{}c@{}@{}c@{}@{}c@{}}
\includegraphics[angle=0,width=0.32\linewidth]{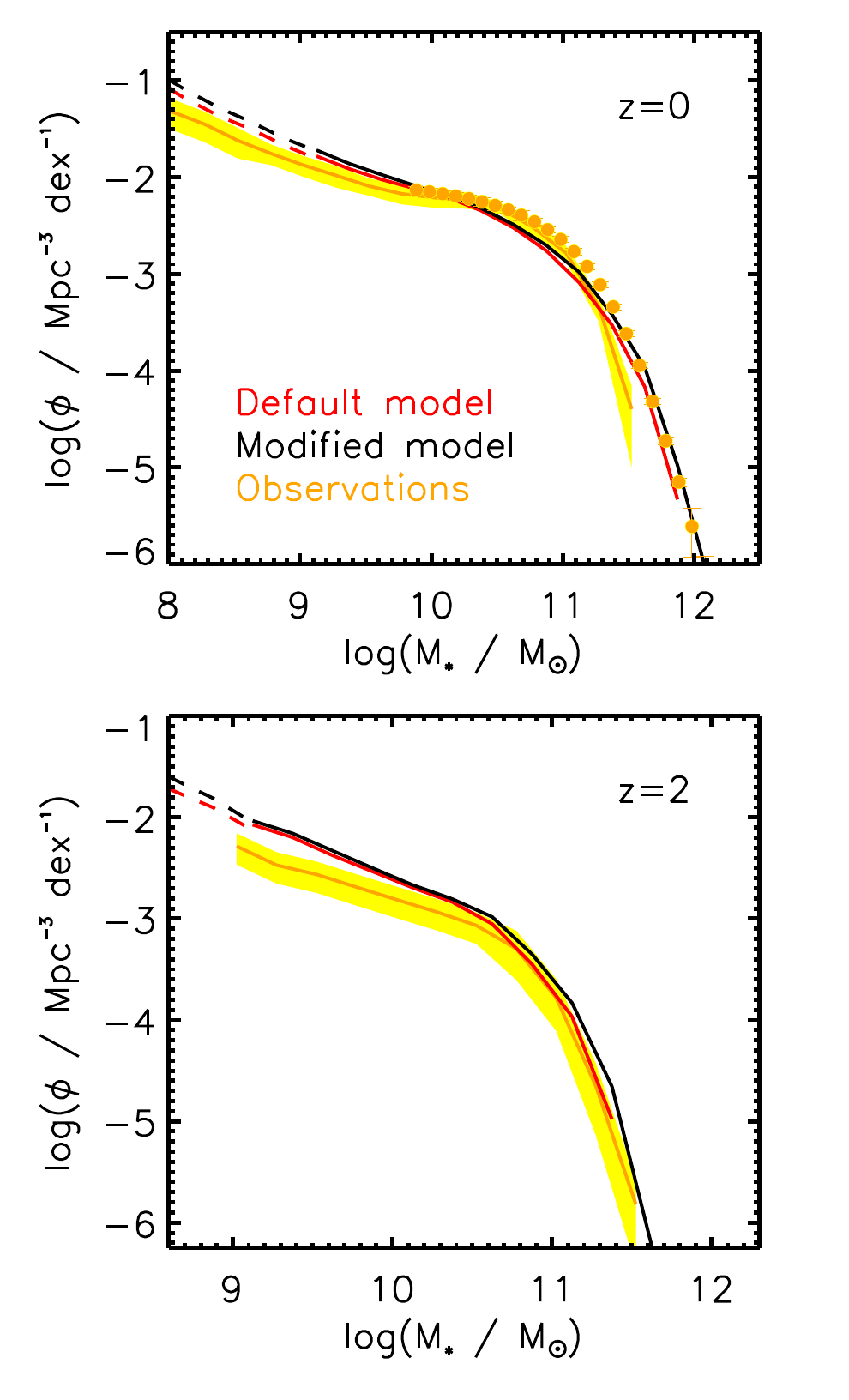} &
\includegraphics[angle=0,width=0.32\linewidth]{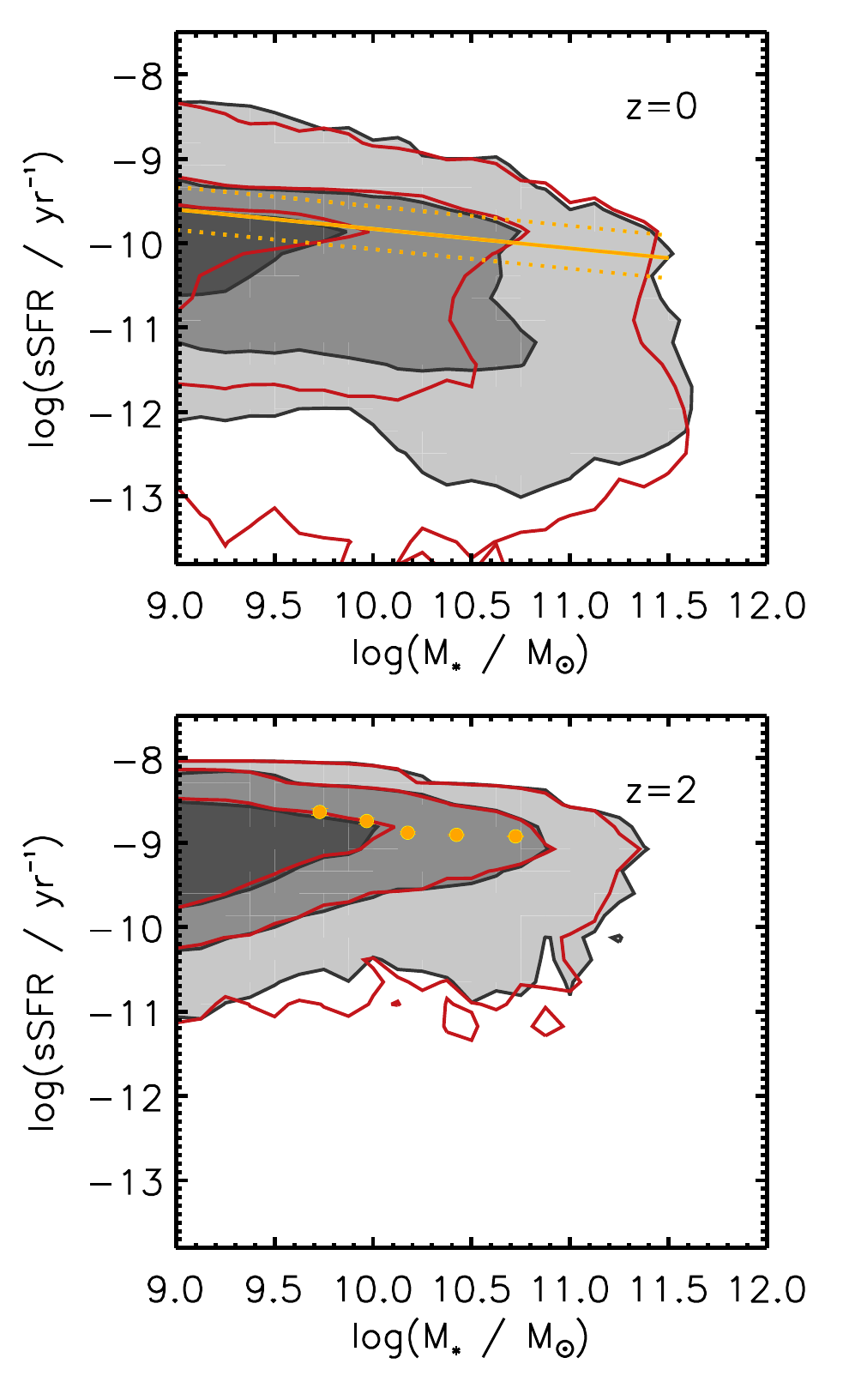} &
\includegraphics[angle=0,width=0.32\linewidth]{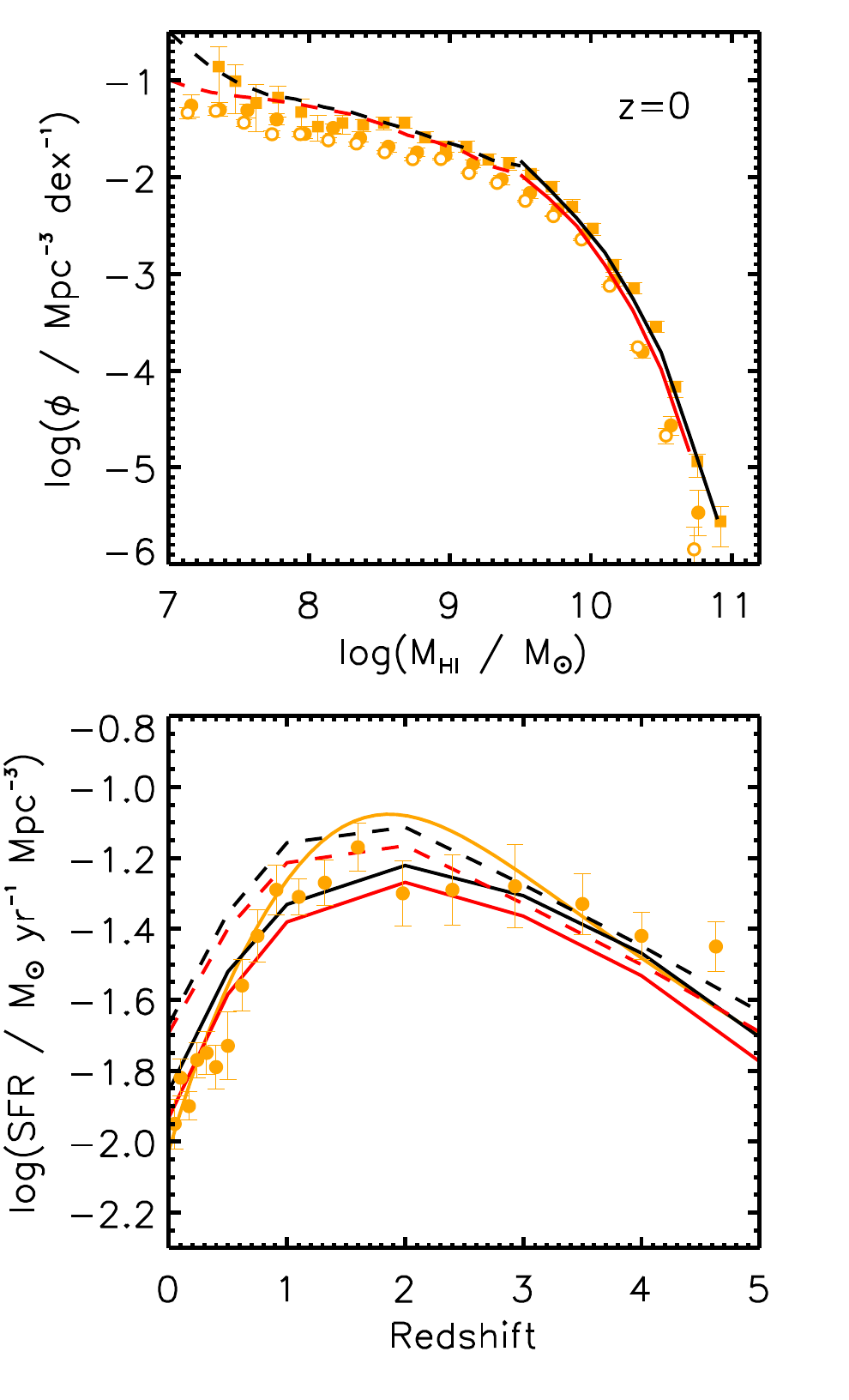}
\end{tabular}
\caption{Key galaxy relations from the \lgaltt{} model. In all panels, red lines represent the default model, black lines represent the modified model, and orange lines and points represent observational data. Solid and dashed lines indicate models when run on the \textsc{Millennium-I} and \textsc{Millennium-II} simulations, respectively. \textit{Top left panel:} The stellar mass function (SMF) for galaxies at $z=0$. The observational data are from \citet{Baldry+08,Baldry+12,Li&White09} (solid line), and \citeauthor{D'Souza+15} (2015, filled circles). \textit{Bottom left panel:} The SMF at $z=2$. Observations are from the combined datasets discussed in appendix A2 of \citealt{Henriques+15}. \textit{Top centre panel:} The $M_{*}$ - sSFR relation for model galaxies, compared to a fit to star-forming systems from the SDSS-DR4 by \citet{Elbaz+07}. \textit{Bottom centre panel:} The $M_{*}$ - sSFR relation at $z=2$, compared to a dataset of 17 systems from the MOSDEF survey \citep{Shivaei+16}. \textit{Top right panel:} The HI mass function (HIMF) at $z=0$, compared to observational datasets from \citeauthor{Zwaan+05} (2005, squares), \citeauthor{Haynes+11} (2011, filled circles), and \citeauthor{Jones+18} (2018, open circles). \textit{Bottom right panel:} The evolution of the cosmic SFR density (SFRD) from $z=5$ to $z=0$, including observational relations from \citeauthor{Madau&Dickinson14} (2014, orange line) and \citeauthor{Driver+18} (2018, orange circles).}
\label{fig:General_relations}
\end{figure*}

Fig. \ref{fig:General_relations} shows a set of key galaxy relations from the \lgaltt{} model. In all panels, red lines represent the default model, black lines represent the modified model, and orange lines and points represent observational data. Solid lines show results when \lgaltt{} is run on the \textsc{Millennium-I} simulation, which produces well-resolved galaxies down to stellr masses of $\logMm{}\sim{}9.0$. Dashed lines show results when run on the higher-resolution \textsc{Millennium-II} simulation \citep{Boylan-Kolchin+09}, which allows us to more accurately probe lower-mass systems.

We can see from the left-hand panels that both versions of \lgaltt{} have very similar stellar mass functions (SMFs), reproducing the knee of the observed SMF at $z=0$ and 2 reasonably well. At the high-mass end, both models are also in good agreement with the observed SMF measured by \citet{D'Souza+15} at $z=0$, which accounts for the faint stellar light in the outskirts of massive elliptical galaxies and is therefore a closer match to the total stellar masses provided by \lgaltt{}. We note that \lgaltt{} is not calibrated to this SMF, but rather to the combined SMFs at $z=0-3$ shown by the solid orange lines (see \citealt{Henriques+15}, appendix A2).


The centre panels show the relation between stellar mass and specific star-formation rate (sSFR $= \tn{SFR}/M_{*}$) for model galaxies at  $z=0$ and 2. They are illustrative of our finding that the `main sequence' of star-forming galaxies is very similar in the default and modified models at all redshifts. Likewise, the top right panel shows the HI mass function (HIMF) at $z=0$, which is also similar in both models. This is because star-forming disc galaxies are able to compensate for the decreased reheating in the modified model by reducing their cooling rates, as these are predominantly determined by the amount of hot gas available. Such systems therefore maintain similar SFRs and total mass return rates in both models, reflecting the self-regulating nature of secularly-evolving, star-forming galaxies seen in equilibrium models (\eg{}\citealt{Bouche+10,Dave+12,Lilly+13}).

The top-centre panel in Fig. \ref{fig:General_relations} also shows a reduced spread of galaxies down to low sSFR in the modified model at low redshift, indicating higher overall SFRs in `red sequence' galaxies (\ie{}ETGs). This is also reflected in the cosmic star-formation rate density (SFRD), shown in the bottom-right panel, where the modified model exhibits a slightly higher SFRD than the default model. This change is due to slightly extended star-formation timescales in ETGs, caused by the reduced mass-loading factors and radial inflow speeds in the modified model compared to the default model. These changes bring the modified model into slightly better agreement with SFRD observations at high redshift, but slightly worse agreement at $z\sim{}0$.

We can also see from the bottom-right panel that the SFRD evolution in both versions of \lgaltt{} is higher when run on \textsc{Millennium-II} than \textsc{Millennium-I}. As already discussed by \citet{Henriques+20}, this is due to a combination of (a) the larger number of very-low-mass galaxies and (b) the higher typical SFRs in these systems when \lgaltt{} is run on the higher-resolution \textsc{Millennium-II} simulation. We note that, with the exception of Section \ref{sec:Profiles evo}, only \textsc{Millennium-I} runs are considered in the rest of this work, as we are interested in assessing the match between \lgaltt{} and observations of disc galaxies above $\logMm{}\sim{}9.0$.

We therefore conclude that the changes made in the modified \lgaltt{} model have a relatively negligible effect on the general global properties of galaxies, particularly the star-forming disc galaxies studied in this work. Given this, a full re-calibration of the model is not required, although we note that such a re-calibration, including the inclusion of the GCE parameters into the MCMC formalism (alongside a robust set of observational metallicity constraints), will be the focus of future work.

\subsection{Model star-forming galaxy sample}\label{sec:Model sample selection}
In order to compare \lgaltt{} to the latest observational IFU data on the distribution of metals across disc galaxies (see Section \ref{sec:MaNGA}), we require a sample of model systems that reflects the same mass range, morphology, and star-formation activity. Therefore, our model galaxy sample is formed by selecting systems with \logM $> 9.0$, $M\sub{disc}/(M\sub{disc}+M\sub{bulge}) > 0.7$, and log(sSFR$/\tn{yr}^{-1}$) $\geq \logten{}[2\cdot{}(1+z)^{2}/(t\sub{H0}/\tn{yr})]-1$, where $t\sub{H0}$ is the Hubble time at $z=0$. This last constraint selects galaxies with log(sSFR$/\tn{yr}^{-1}$) $\gtrsim -10.9$ at $z=0$, increasing to $\gtrsim -9.7$ at $z=3$ (see \citealt{Henriques+20}, appendix A), and is found to accurately select the star-forming main sequence in both the default and modified model back to at least $z=7$. This leaves us with $\sim{}$80,000 galaxies at $z=0$ in our star-forming model sample.

\section{The MaNGA sample}\label{sec:MaNGA}
In this work, we predominantly compare the low-redshift chemical properties of model galaxies to observational data from the Mapping Nearby Galaxies at the Apache Point Observatory (MaNGA) survey \citep{Bundy+15}. MaNGA obtains spatially-resolved maps out to $1.5-2.5$ effective radii ($R\sub{e}$), by utilising the IFU and spectrograph mounted on the Sloan 2.5m optical telescope. Each MaNGA galaxy datacube is made up of spectra from a bundle of 19-127 fibres, providing a field-of-view of 12 - 32 arcsec (equivalent to 9.3 - 24.7 kpc at the median redshift of our MaNGA sample, $z\sim{}0.037$).

Our MaNGA sample is drawn from SDSS data release (DR) 15. We utilise the derived data products from the MaNGA Data Analysis Pipeline (DAP) presented by \citet{Westfall+19} and \citet{Belfiore+19}. This gives us access to Gaussian-profile integrated optical emission line fluxes and ancillary data for a base sample of 4,648 galaxies. We cross-match this base sample with the MaNGA FIREFLY v2.4.3 value-added catalogue (VAC) of 4,605 DR15 galaxies \citep{Wilkinson+17,Goddard+17}. This VAC provides absorption-line based stellar metallicities and ages, obtained using the FIREFLY spectral-fitting code which utilises the stellar population synthesis models of \citet{Maraston&Stromback11}. This combination of emission-line and absorption-line data allows us to study both the gas and stellar properties in our sample galaxies from the same spectra. We take stellar masses from the NASA Sloan Atlas (NSA) v1.0.1\footnote{\textit{https://www.sdss.org/dr13/manga/manga-target-selection/nsa/}} \citep{Blanton+11}, which are fit to elliptical Petrosian fluxes. We have converted these stellar masses to our assumed cosmology (\ie{}$h=0.68$) by multiplying by a factor of $1/h^{2}\sim{}2.16$.

\subsection{MaNGA sample selection}\label{sec:Sample selection}
The MaNGA DAP provides emission-line information in two formats; individual spaxel spectra (the HYB10 datacubes), and co-added spectra for groups of spaxels  (the VOR10 datacubes). When studying gas-phase (\ie{}\HII{} region) metallicities in MaNGA, we utilise the HYB10 spectra. This allows us to take advantage of their higher spatial resolution, noting that lower spatial resolution spectra can return erroneously flatten radial metallicity gradients (see \eg{}\citealt{Poetrodjojo+19,Acharyya+20}). Because we are predominantly interested in azimuthally-averaged radial profiles and are measuring line ratios rather than absolute line fluxes, the leakage of light across HYB10 spaxels due to the large point spread function (PSF) in MaNGA should not be a significant issue (although, see Section \ref{sec:Metallicity diagnostics}).

In order to make a fair analysis of the radial metallicity profiles within galaxy discs, we restrict our MaNGA sample to systems with an inclination angle\footnote{Following \citet{Giovanelli+94}, we assume an `intrinsic ellipticity' (caused by an intrinsic disc thickness) of $q\sub{el}=0.13$ when calculating the inclination angle from the observed axial ratios ($b/a$) provided by the NSA catalogue, where $\tn{cos}^{2}i=[(b/a)^{2}-q\sub{el}^{2}]/(1-q\sub{el}^{2})$.} of $0^{\circ} \leq i < 60^{\circ}$, and which are identified as spiral galaxies in the Galaxy Zoo DR1 catalogue \citep{Lintott+08,Lintott+11} with a vote fraction of $>80$ per cent. This reduces our sample to 663 systems, with the greatest reduction coming from the spiral morphology requirement, which is found to hold for only 31 per cent of the full FIREFLY sample. This morphological requirement can have an important effect on the average \Zgt{} profiles obtained for low-mass systems (see Section \ref{sec:Mass dependent Zg profiles}).

We also impose a cut on the global specific star-formation rate, selecting only galaxies with log(sSFR/yr$^{1}$) $\geq -11$, within errors. For this we utilised the total SFRs provided by the SDSS-DR7 catalogue (see \citealt{Brinchmann+04,Salim+07}) and the stellar masses provided by the NSA catalogue. We also only consider datacubes with a MaNGA DAP quality flag \texttt{DAPQUAL} of 0 (see \citealt{Westfall+19}, table 10) and spaxels/cells which fall within the MaNGA DAP quality mask.

For our \Zgt{} analysis based on optical emission lines, we also require a minimum S/N of 3 on all lines used, and an equivalent width in the H$\alpha$ line of EW(\Ha) $> 14$\AA{}. This latter criterion minimises the contamination by diffuse ionised gas (DIG, see \eg{}\citealt{Sanders+17,Lacerda+18}). We also only select spectra which fall below the empirical demarcation line provided by \citet{Kauffmann+03a} for the \NII{}/\Ha{} -- \OIII{}/\Hb{} BPT diagram \citep{Baldwin+81}. This removes regions for which the ionising source is unlikely to be young stars, although we note that the clear majority of these regions are already removed by the EW(\Ha) $> 14$\AA{} cut. Finally, we correct all emission-line fluxes for internal dust extinction via the \Ha{}/\Hb{} ratio, using an intrinsic Balmer decrement of 2.86 and the \citet{Calzetti+00} attenuation law for star-forming galaxies, which assumes an extinction factor of $R^{'}\sub{V}=4.05\error{0.08}$. We have checked that small changes to these selection criteria do not significantly affect our results. For example, while holding all other criteria fixed, varying the emission-line S/N threshold between 0 and 3, varying the EW(\Ha) threshold between 3 and 14\AA{}, and varying the maximum permitted inclination between 40 and $60^{\circ}$, does not alter the average metallicity profiles we obtain for each stellar mass bin.


\begin{figure*}
\centering
\includegraphics[angle=0,width=0.99\linewidth]{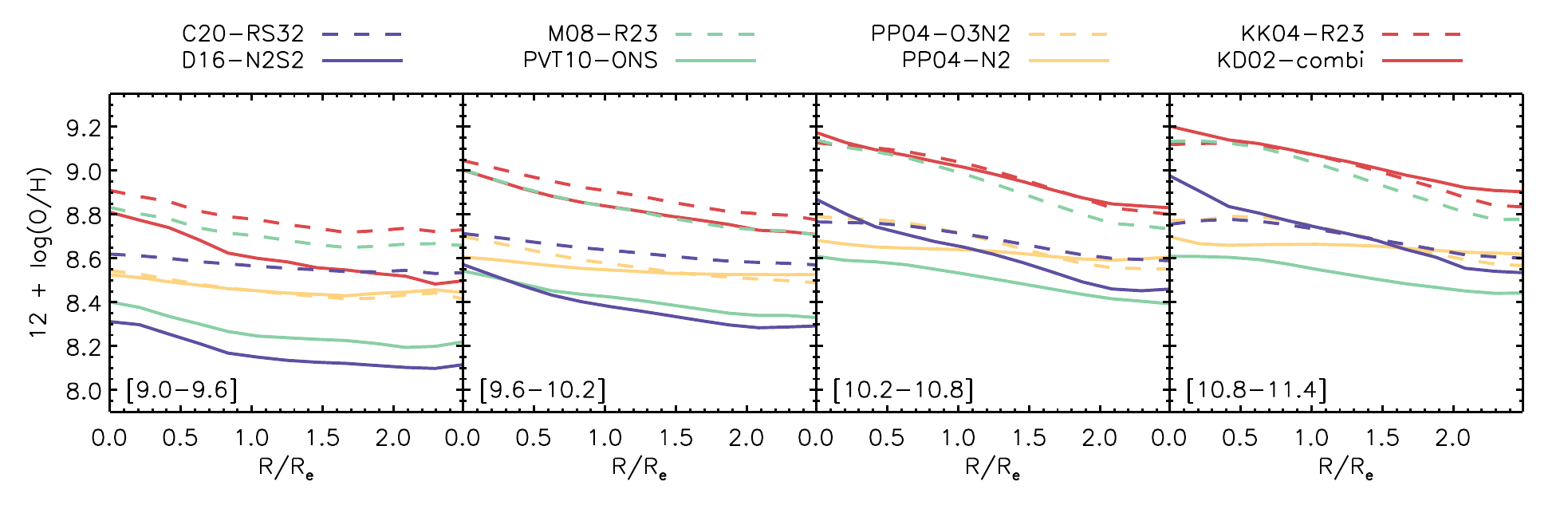}
\caption{Gas-phase metallicity (\Zgt{}) profiles as a function of $R/R\sub{e}$ from our MaNGA sample, split into four bins of stellar mass. \Zgt{} is calculated using the following eight strong line (SL) metallicity diagnostics from the literature: Dashed blue lines, \OIII{}/\Hb{} + \SII{}/\Ha{} from \citeauthor{Curti+20} (2020, C20-RS32); solid blue lines, \NII{}/\SII{} + \NII{}/\Ha{} from \citeauthor{Dopita+16} (2016, D16-N2S2); dashed green lines, (\OIII{}+\OII{})/\Hb{} from \citeauthor{Maiolino+08} (2008, M08-R23); solid green lines the oxygen-nitrogen-sulphur combined diagnostic from \citeauthor{Pilyugin+10b} (2010, PVT10-ONS); dashed orange lines, \OIII{}/\NII{} from \citeauthor{Pettini&Pagel04} (2004, PP04-O3N2); solid orange lines, \NII{}/\Ha{} from \citeauthor{Pettini&Pagel04} (2004, PP04-N2); dashed red lines, (\OIII{}+\OII{})/\Hb{} from \citeauthor{Kobulnicky&Kewley04} (2004, KK04-R23) which includes an iterative $q$ correction using \OIII{}/\OII{}; and solid red lines, the combined diagnostic proposed by \citeauthor{Kewley&Dopita02} (2002, KD02-combi) which utilises \NII{}/\OII{} for metallicities above 8.6 dex.}
\label{fig:Zprofiles_SLdiags}
\end{figure*}

The \Zst{} estimates calculated by FIREFLY utilise the VOR10 datacubes, which are formed by binning spaxels into `spatial cells' to obtain S/N $\geq 10$ in the $g$-band continuum \citep{Cappellari&Copin03}. For our \Zst{} analysis, we select those VOR10 spatial cells with an uncertainty in their luminosity-weighted \Zst{} of less that 0.23 dex. This corresponds to an average S/N in the r-band of 5 (see \citealt{Goddard+17}, section 2.2). Following \citealt{Goddard+17}, we also only consider spatial cells within 1.5 effective radii (\Reff{}), using the galactocentric radii normalised to the ellipitical Petrosian r-band effective radius provided by the NSA catalogue.

These selection criteria leave us with a final MaNGA sample of 571 near-face-on disc galaxies, containing 450,635 spaxels with measurable \Zgt{} (when using our preferred metallicity diagnostic, see below), and 272,975 spatial cells with measured \Zst{} from the FIREFLY catalogue.

\subsection{Metallicity diagnostics}\label{sec:Metallicity diagnostics}
By far the biggest influence on the \Zgt{} profiles we derive for our MaNGA sample comes from the choice of strong-line (SL) metallicity diagnostic. It is already well established that different SL diagnostics return a wide range of \Zgt{} estimates, varying by up to 0.7 dex for the same spectra \citep{Kewley&Ellison08}. This has a significant impact on the scaling relations derived from samples of global spectra (\eg{}\citealt{Yates+12}), and also affects the interpretation of galaxy radial metallicity profiles (\eg{}\citealt{Maiolino&Mannucci19}).

Fig. \ref{fig:Zprofiles_SLdiags} shows the mean radial \Zgt{} profiles we obtain for our MaNGA sample when stacked by $R/R\sub{e}$, for four bins of stellar mass. Eight different SL \Zgt{} diagnostics from the literature are considered: KD02-combi \citep{Kewley&Dopita02}, KK04-R23 \citep{Kobulnicky&Kewley04}, PP04-N2 \& PP04-O3N2 \citep{Pettini&Pagel04}, M08-R23 \citep{Maiolino+08}, PVT10-ONS \citep{Pilyugin+10b}, D16-N2S2 \citep{Dopita+16}, and C20-RS32 \citep{Curti+20}. These diagnostics have been chosen to represent a range of different emission-line ratios and calibration samples.

The first noticeable feature in Fig. \ref{fig:Zprofiles_SLdiags} is the large difference in normalisation between the profiles at fixed mass. The variation of up to $\sim{}0.6$ dex seen here is reminiscent of that seen for the global M\Zgt{}R when using different SL diagnsotics \citep{Kewley&Ellison08}, and is arguably more significant than the variation in slope (although see Section \ref{sec:Mass dependent Zg profiles}). The second noticeable feature is the large difference in mass dependence between the different SL diagnostics. Some suggest a large increase in \Zgt{} profile normalisation with stellar mass (\eg{}KD02-combi, D16-N2S2), whereas others suggest essentially no change in metallicity with mass at all (\eg{}PP04-N2, C20-RS32).

As shown by \citealt{Yates+20} and others, SL diagnostics which return a low \Zgt{} at low mass (\eg{}PVT10-ONS and D16-N2S2) are in better agreement with metallicities obtained from a variety of direct methods, such as metal recombination lines \citep{Esteban+09,Esteban+14}, absorption lines from blue supergiant photospheres \citep{Kudritzki+16}, and electron temperature ($T\sub{e}$) measurements (\eg{}\citealt{,Bresolin+09b}). Such methods are typically considered more accurate than SL diagnostics in low-metallicity environments (\eg{}\citealt{Bresolin08}). The other SL diagnostics considered here, particularly KD02-combi and M08-R23, predict global metallcities at low mass which are higher than expected from direct measurements by up to $\sim{}0.45$ dex.

At higher masses, the PVT10-ONS diagnostic returns particularly low \Zgt{} estimates, predicting that star-forming disc galaxies with \logM{} $\sim{}11.5$ should have sub-solar ISM metallicities at all radii. Similarly, the PP04-N2 diagnostic is known to saturate at around solar metallicity, prohibiting its use for the highest-mass systems (see \eg{}\citealt{Kewley&Dopita02}). On the other hand, the D16-N2S2 diagnostic returns high \Zgt{} at high mass, predicting super-solar ISM metallicities within $\sim{}1$ \Reff{} at Milky-Way masses and above.

Another interesting feature seen in Fig. \ref{fig:Zprofiles_SLdiags} is the differences in inner slope exhibited by different SL diagnostics for massive galaxies (see also \citealt{Boardman+20}). Those diagnostics which rely on \OIII{} lines (\ie{}KK04-R23, PP04-O3N2, M08-R23, PVT10-ONS, C20-RS32) all suggest a flattening within $\sim{}0.5\,R\sub{e}$ at log$(M_{*}/\Msun) \geq 10.2$ (\ie{}the two highest-mass bins), whereas those which do not (\ie{}KD02-combi, PP04-N2, D16-N2S2) suggest a continued increase in metallicity towards the centre.

\begin{figure*}
\centering
\includegraphics[angle=0,width=0.99\linewidth]{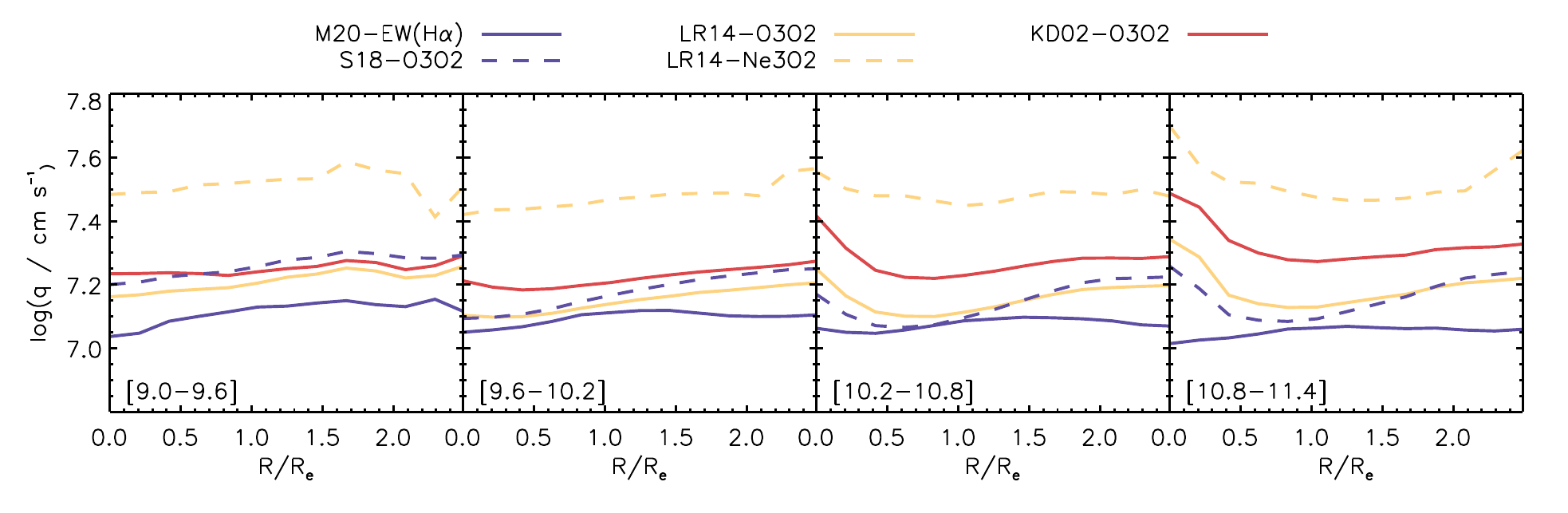}
\caption{Ionisation parameter ($q$) profiles as a function of $R/R\sub{e}$ for our MaNGA sample, split into four mass bins. Profiles are obtained using the following five SL $q$ diagnostics: M20-EW(H$\alpha$) (\citealt{Mingozzi+20}, solid blue lines), S18-O3O2 (\citealt{Strom+18}, dashed blue lines), LR14-O3O2 (\citealt{Levesque&Richardson14}, solid orange lines), LR14-Ne3O2 (\citealt{Levesque&Richardson14}, dashed orange lines), KD02-O3O2 (\citealt{Kewley&Dopita02}, solid red lines).}
\label{fig:Uprofiles}
\end{figure*}

We note that \citet{Belfiore+17} caution against drawing physical conclusions from metallicity gradients within $\sim{}0.5R\sub{e}$ in MaNGA, due to the beam-smearing effects caused by high PSF/$R\sub{e}$ values. This can lead to a slight but systematic flattening of measured \Zgt{} profiles in the inner regions of galaxies. Inclination effects can also contribute in this respect \citep{Belfiore+17}, and physical effects such as radial gas motions could also play a role \citep{Sanchez+14}. Nonetheless, the fact that the SL diagnostics considered here exhibit strong and \textit{diverse} changes to the inner slope suggests that diagnostic-specific biases should also be considered.


A common candidate for such SL diagnostic biases is a dependence on the ionisation state of the line-emitting gas. This is typically represented by the ionisation parameter, $q = S(\tn{H}^{0})/n\sub{\textsc{H}}$, which is the ratio of the ionising photon flux density to the electron density and describes the speed of propagation of an ionising front through hydrogen. Fig. \ref{fig:Uprofiles} shows the radial $q$ profiles for our MaNGA sample, using five different SL $q$ diagnostics. The \OIII{}/\OII{} and \NeIII{}/\OII{} ratios used for most of these are themselves dependent on metallicity, so we have corrected them using the SL \Zgt{} diagnostics described above. We find that the $q$ profiles obtained are the same regardless of which SL \Zgt{} diagnostic is used for this correction.

Fig. \ref{fig:Uprofiles} shows evidence of a clear increase in $q$ towards the centres of massive galaxies. This can affect the \Zgt{} profiles returned by SL \Zgt{} diagnostics which have a secondary dependence on $q$. For example, the O3N2 ratio is known to have a strong positive dependence on $q$ (\eg{}\citealt{Kewley&Dopita02}). Therefore, O3N2 diagnostics calibrated to samples exhibiting a strong one-to-one anti-correlation between $q$ and \Zgt{} (likely the case for most empirical \HII{} region samples) would under-predict the true metallicity for high-$q$ regions. This could partly explain the flattened inner profile returned by diagnostics such as PP04-O3N2 in Fig. \ref{fig:Zprofiles_SLdiags}. However, this is not necessarily the case for other SL diagnostics which return flattened inner profiles. For example, the KK04-R23 diagnostic, which has a negative $q$ dependence and iteratively corrects for this while solving for \Zgt{}, also returns a flattened profile at low radii in massive galaxies.

We therefore conclude that $q$ dependencies could affect the observed inner \Zgt{} profile in high-mass galaxies, but the nature of this affect is complex and likely dependent on the line ratio and calibration sample used. Consequently, we recommend the use of SL diagnostics which do not have a strong $q$ dependence, in order to avoid such issues. Further investigation using metallicity-independent $q$ diagnostics, direct measurements of \Zgt{}, and a wider range of spatial scales is required in order to draw more comprehensive conclusions (see Easeman et al., in prep.).

Given the arguments above, in this work we choose to adopt the D16-N2S2 diagnostic when estimating MaNGA gas-phase metallicities in this work. This diagnostic is well matched in normalisation to direct metallicity measurements at low mass (see Section \ref{sec:Zg}), predicts super-solar \Zgt{} in very massive galaxies, has a relatively negligible dependence on $q$, and is also calibrated assuming the observed O/H -- N/O relation obtained from direct measurements of stellar and \HII{}-region spectra (see \citeauthor{Dopita+16} 2016, section 4.2). This last factor is important given the use of the \NII{}$\lambda$6584 line in this diagnostic when estimating the oxygen abundance.

Nonetheless, we note that, like the SL ratios used in some other common diagnostics (\eg{}N2 and O3N2), the \NII{}/\SII{} and \NII{}/\Ha{} ratios used in the D16-N2S2 diagnostic do not directly contain a measurement of collisionally-excited oxygen lines to hydrogen recombination lines. This diagnostic instead relies on a good correspondence between the assumed O/H -- N/O relation and that of the particular sample studied. The position of a galaxy on the O/H -- N/O relation can depend on its $M_{*}$ and SFR (\eg{}\citealt{Andrews&Martini13}), and more generally on the evolutionary stage of the galaxy (\eg{}\citealt{Vincenzo+16}). In our case, the O/H -- N/O relation for our MaNGA sample is relatively tight and its slope is in good qualitative agreement with that expected by \citet{Dopita+16}, when considering a number of different SL N/O and O/H diagnostics. The metallicity profiles we present are also averages in bins of stellar mass, and we note that \citet{Andrews&Martini13} find galaxies stacked by mass return a tight and consistent O/H-N/O trend. Therefore, we do not expect biases due to deviations in the star formation histories among our sample to be significant here.

\section{Results}\label{sec:Results}
Before discussing radial profiles, we first present the global metallicities present in the ISM, stars, CGM, and ICM at low redshift, and discuss their evolution over cosmic time.


\begin{figure}
\centering
 \includegraphics[angle=0,width=0.99\linewidth]{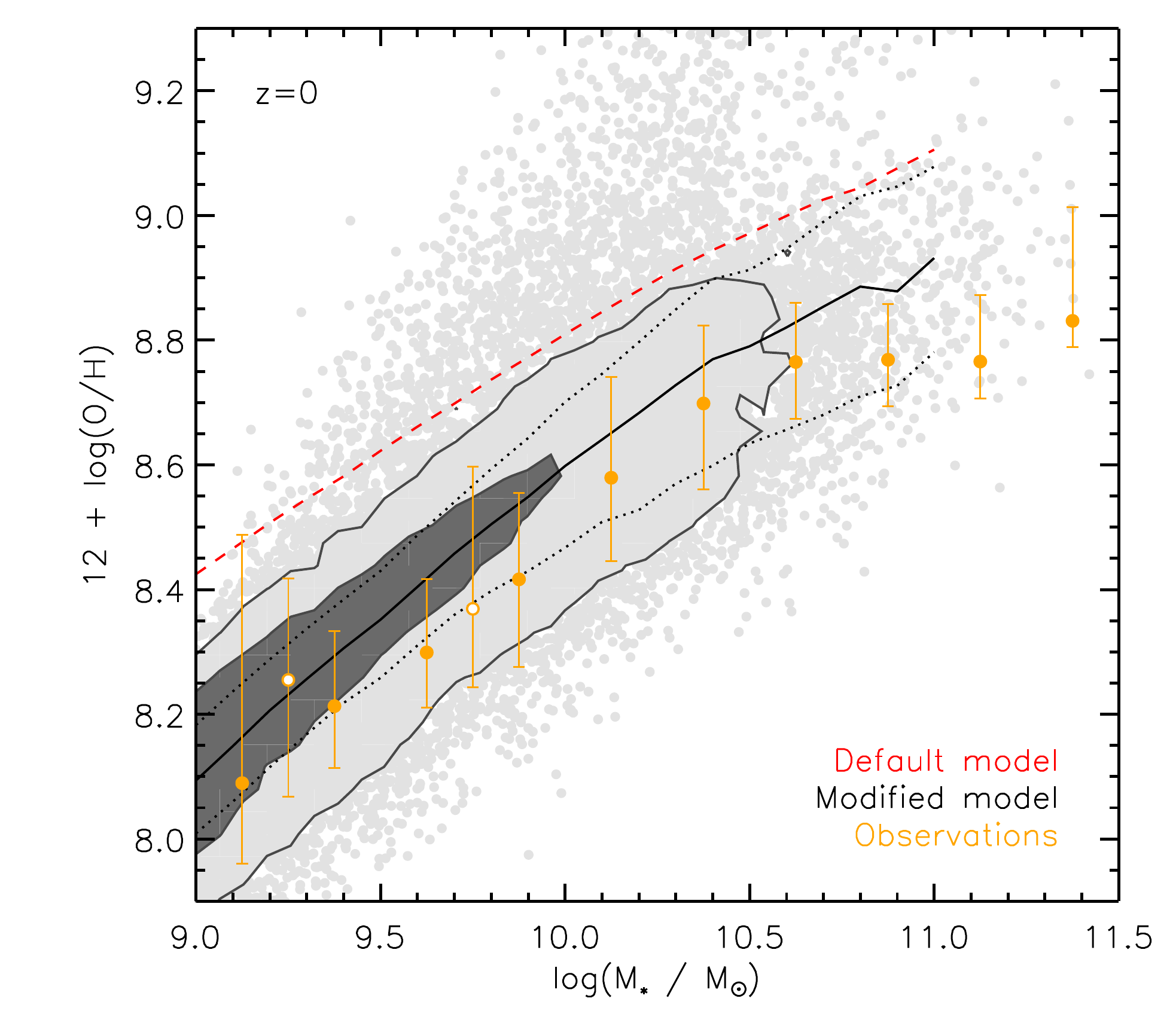}
 \caption{The relation between stellar mass and SFR-weighted gas-phase metallicity (the M\Zgt{}R) for star-forming galaxies at $z=0$. Contours denote the $1\sigma$ and $2\sigma$ distributions for the modified \lgaltt{} model, the solid black line denotes the mean relation, and the dotted black lines denote the 16th and 84th percentiles. The mean relation from the default model is shown as a red dashed line. The two observed binned M\Zgt{}Rs also shown are from \Te{}-based metallicity measurements made by \citeauthor{Yates+20} (2020, open orange circles), and strong-line metallicities using the D16-N2S2 diagnostic for the MaNGA sample presented in Section \ref{sec:MaNGA} (orange circles).}
 \label{fig:MZgR_z0}
\end{figure}

\subsection{Global metallicities at low redshift}\label{sec:MZRs at z0}

\subsubsection{ISM metallicities at low redshift}\label{sec:Zg}
Fig. \ref{fig:MZgR_z0} shows the relation between stellar mass and ISM metallicity (the M\Zgt{}R) at $z=0$ for star-forming, disc-dominant galaxies. Model galaxies from the modified model are represented by grey contours and points. The mean relation for the modified model (black, solid line) and the default model (red, dashed line) are also shown. In order to best mimic observations, global \Zgt{} in \lgaltt{} is obtained by first calculating the local \Zgt{} within each radial ring in units of 12+log(O/H),\footnote{Where O/H is the ratio of the number density of oxygen atoms to hydrogen atoms in the gas.} and then taking the SFR-weighted mean of these local values.

The observational M\Zgt{}R from our MaNGA sample is also shown as filled orange circles, with the vertical error bars representing the 1$\sigma$ spread in \Zgt{} in each mass bin. This is formed by calculating the global H$\alpha$-flux-weighted \Zgt{} (using the D16-N2S2 SL diagnostic, see Section \ref{sec:Metallicity diagnostics}) for each MaNGA galaxy, and plotting the mean of these global metallicities in 0.25 dex wide bins of stellar mass. The upper end of the M\Zgt{}R, derived from galaxies with \Te{}-based \Zgt{} measurements \citep{Yates+20}, is also shown (open orange circles). We can see a good agreement between the D16-N2S2-based and \Te{}-based M\Zgt{}Rs, as discussed in Section \ref{sec:Metallicity diagnostics}.

\begin{figure}
\centering
 \includegraphics[angle=0,width=0.99\linewidth]{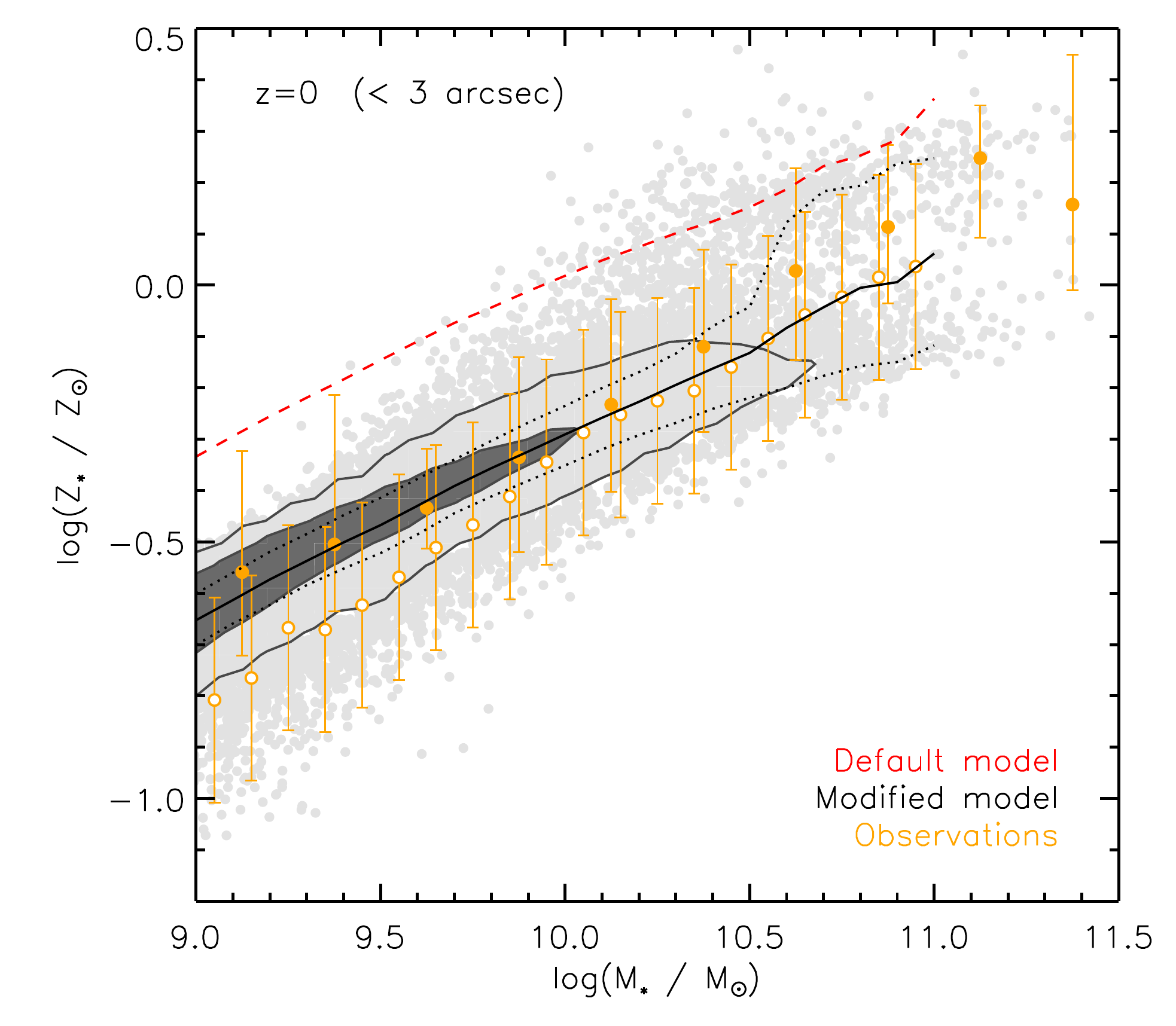}
 \caption{The relation between stellar mass and mass-weighted stellar metallicity (the M\Zst{}R) within 3 arcsec for star-forming galaxies at $z=0$. Lines, contours, and colours are as in Fig. \ref{fig:MZgR_z0}. The two observed M\Zst{}Rs shown are from \citeauthor{Zahid+17} (2017, open orange circles) and our MaNGA sample presented in Section \ref{sec:MaNGA} (filled orange circles).}
 \label{fig:MZsR_z0}
\end{figure}

The M\Zgt{}R from the modified model agrees well with the observations shown in Fig. \ref{fig:MZgR_z0}. The decrease in normalisation compared to the default model is due to the reduction in metal enrichment of the ISM that occurs when the $f\sub{hot}$ parameters are increased.

We also find that, at $\logMm{}\lesssim{}10.2$, both the default and modified models reproduce an anti-correlation between SFR and \Zgt{} at fixed mass, as expected from the fundamental metallicity relation (FMR, \citealt{Mannucci+10}). At higher mass, this SFR-\Zgt{} trend is reversed in the modified model (as seen in previous versions of \lgal{} and some observations, \citealt{Yates+12}), but not in the default model. This difference is likely due to the more significant impact that increased metal removal has on low-SFR massive galaxies in our model gaalxy sample. This is an interesting finding which we intend to investigate further in future work.

We note that the default \lgaltt{} model better matches the M\Zgt{}Rs inferred from theoretically-calibrated \OIII{}-based SL diagnostics, such as those from \citet{Kewley&Dopita02} and \citet{Maiolino+08} (see \citealt{Henriques+20}, fig. 13). However, unlike the D16-N2S2 diagnostic chosen in this work, those SL diagnostics are inconsistent with direct \Zgt{} measurements for low-mass galaxies and \HII{} regions (see Section \ref{sec:Metallicity diagnostics}). Higher ISM metallicities in the default model also lead to a much reduced evolution in the M\Zgt{}R back to high redshift, which is discussed further in Section \ref{sec:Z_evo}.

\subsubsection{Stellar metallicities at low redshift}\label{sec:Zs}
Fig. \ref{fig:MZsR_z0} shows the relation between stellar mass and stellar metallicity (the M\Zst{}R) at $z=0$ for star-forming, disc-dominant galaxies. To calculate global stellar metallicities in \lgaltt{}, we first measure the solar-normalised metal-to-total stellar mass ratio in each radial ring, log$(\Zs{}/\Zsun) = \logten{}(M_{*,\tn{Z}}/M_{*}) - \logten{}(\Zsun)$, where $\Zsun = 0.0142$ is the metallicity in the bulk of the Sun \citep{Asplund+09}, and then take the stellar-mass-weighted mean as the global \Zst{}.

The observational M\Zst{}Rs for low-redshift galaxies plotted in Fig. \ref{fig:MZsR_z0} come from two sources, (a) mass-weighted, central \Zst{} measurements from our MaNGA sample provided by the FIREFLY catalogue (filled orange circles), and (b) mass-weighted, central \Zst{} measurements from absorption lines in stacked spectra of $\sim{}200,000$ SDSS-DR7 star-forming galaxies from \citet{Zahid+17} (open orange circles). We have assigned a nominal uncertainty of 0.2 dex to the \citet{Zahid+17} binned data, given that their complimentary luminosity-weighted metallicities are expected to be accurate to within $\sim{}0.1$ dex, with the uncertainty in the mass-weighted counterparts being larger due to additional dependencies on the assumed star formation history (H. J. Zahid, priv. comm.).

All observational \Zst{} measurements used in Fig. \ref{fig:MZsR_z0} have been obtained within a 3 arcsec aperture -- the diameter of an SDSS fibre. This equates to 2.29 kpc at the median redshift of our MaNGA sample ($\tilde{z} = 0.037$). At low redshifts, such limited apertures can return significantly higher \Zst{} estimates, due to the presence of negative metallicity gradients in galaxies (see \eg{}\citealt{Okamoto+17}). Therefore, we measure global \Zst{} for our model galaxies only within the innermost 7 rings for Fig. \ref{fig:MZsR_z0}, equating to a diameter of 1.88 kpc, in order to better compare with observations. This increases the mean \Zst{} by $\sim{}0.12$ dex at all masses in the model. Interestingly, this aperture `anti-correction' also reduces the 1$\sigma$ spread in \Zst{} by $\sim{}0.05$ dex at low mass, indicating that variations among galaxies' \Zst{} at larger radii could be an important component of the scatter in the M\Zst{}R.

A key result of this work is that the modified \lgaltt{} model is able to simultaneously reproduce both the M\Zgt{}R and M\Zst{}R at $z=0$. The improved agreement here is due to the improved accuracy and precision of the observational data considered, the choice of \Zgt{} diagnostic used, and the significant decrease in ISM enrichment efficiency in the modified model compared to the default model.

\subsubsection{CGM metallicities at low redshift}\label{sec:Zh}
It is important to also consider the metal content in the hot gas surrounding galaxies, particularly given that our modified model significantly increases the efficiency with which SNe can directly pollute the CGM.

\begin{figure}
\centering
 \includegraphics[angle=0,width=0.99\linewidth]{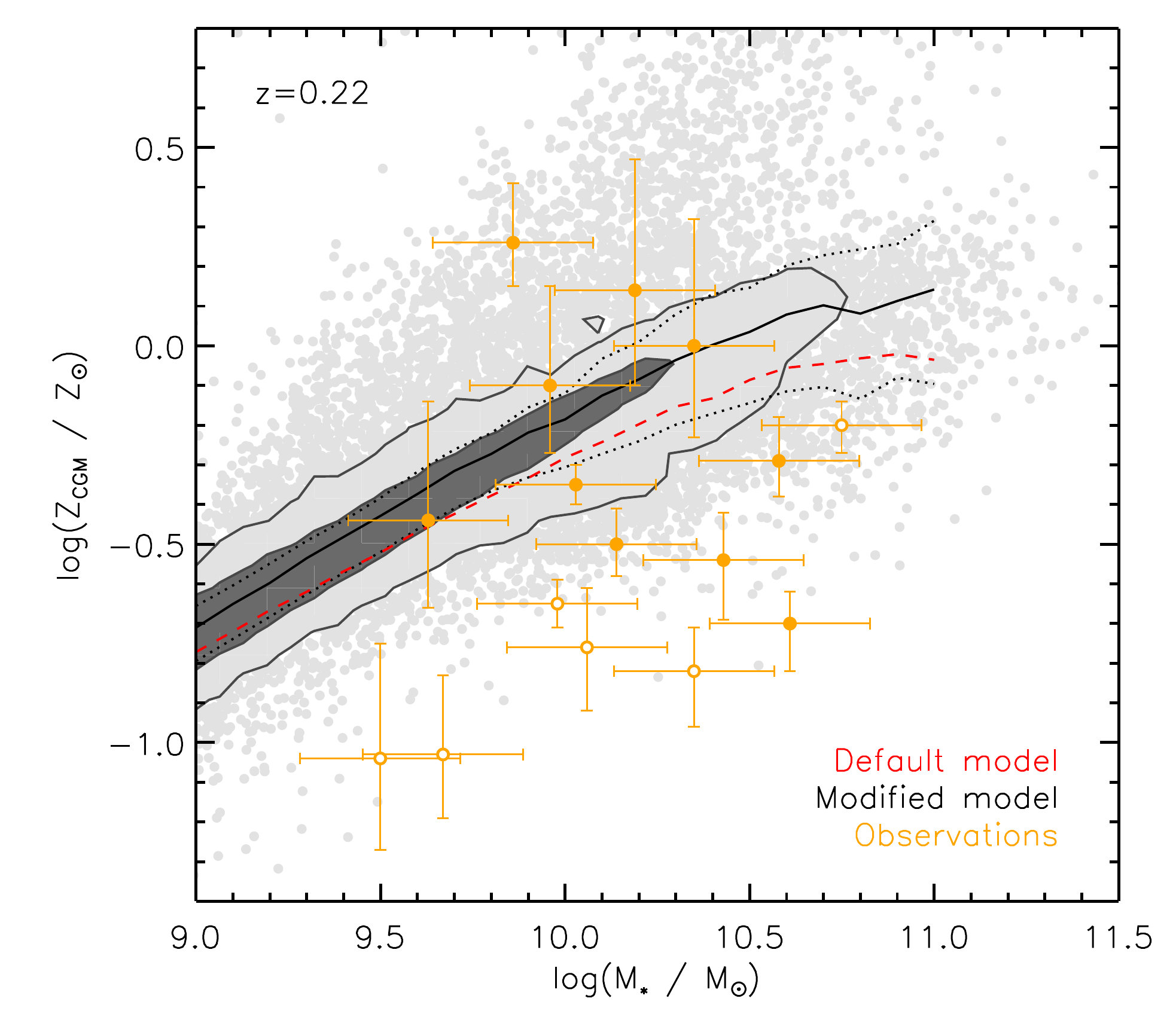}
 \caption{The relation between stellar mass and total metallicity in the CGM (the M\Zht{}R) for star-forming galaxies at $z=0.22$. Lines, contours, and colours are as in Fig. \ref{fig:MZgR_z0}. The observations shown here are taken from the COS-Halos Survey \citep{Tumlinson+11,Prochaska+17}: filled orange circles denote `low-density systems', and open orange circles denote `Lyman limit systems' (LLSs).}
 \label{fig:MZhR_z022}
\end{figure}

Fig. \ref{fig:MZhR_z022} shows the relation between stellar mass and CGM metallicity (excluding ejected material beyond $R\sub{vir}$) for star-forming, disc-dominant galaxies at $z=0.22$. Model metallicities are measured as $\logten{}(\Zh{}/\Zsun) = \logten{}(M\sub{h,Z}/M\sub{h}) - \logten{}(\Zsun)$, where $\Zsun = 0.0134$ is the solar photospheric metallicity from \citet{Asplund+09} assumed in the observations to which we compare here.

The observational data shown in Fig. \ref{fig:MZhR_z022} is taken from the re-analysis by \citet{Prochaska+17} of absorption-line-based metallicities from the COS-Halos Survey \citep{Tumlinson+11,Tumlinson+17}. These metallicities were obtained via Cloudy \citep{Ferland+13} modelling of Si$^{+}$ and Si$^{++}$ absorption lines, assuming solar relative abundances from \citet{Asplund+09} and a \citet{Haardt&Madau12} EUVB radiation field. We have corrected the stellar masses and SFRs from the COS-Halos sample (provided by \citealt{Werk+12}) to assume a \citet{Chabrier03} IMF and our value of $h=0.68$, and have selected only systems with $\logten{}(\tn{sSFR/yr}^{-1}) > -11$ and 1$\sigma$ uncertainties in \Zht{} of $< 0.5$ dex. This leaves 16 systems with $\bar{z}=0.22$ and a median impact parameter of 64 kpc. This observational data set has then been split into two sub-samples: (a) `low-density systems' with $15.0 < \logten{}(N\sub{HI}/\tn{cm}^{-2}) < 17.2$ (filled orange circles) which are optically thin to ionizing radiation and directly trace the hot CGM, and (b) `Lyman limit systems (LLSs)' with $17.2 < \logten{}(N\sub{HI}/\tn{cm}^{-2}) < 19.0$ (open orange circles) which are found in simulations to have neutral gas fractions of $\sim{}10^{-2}$ and typically reside close to galaxies (see \citealt{Peroux&Howk20}, section 1.2). It is the low-density systems which provide the best comparison to the CGM modelled in \lgaltt{}, which is considered to contain hot, non-star-forming gas extending out to the virial radius.

The scatter in the observational data in Fig. \ref{fig:MZhR_z022} is considerably larger than that seen in \lgaltt{}. This could be partly due to variations in the impact parameter on the observational side, and the flat metallicity gradients assumed for the CGM on the model side. However, as also found by \citet{Prochaska+17}, we find no trend in \Zht{} with impact parameter (even at fixed mass) for the COS-Halos data, suggesting that inner metallicity gradients are not systematically present in the CGM of star-forming galaxies.

The mean \Zht{} for the COS-Halos low-density systems is $\Zh{}/\Zsun = -0.25$, which can be compared to $\Zh{}/\Zsun = -0.24$ for the default \lgaltt{} model and $\Zh{}/\Zsun = -0.13$ for the modified model at the same mean stellar mass of \logM{} $\sim{}10.1$. This suggests that there is a slight over-abundance of metals in the CGM for the modified model compared to observations. However, we note that this excess metal is predominantly in light $\alpha$ elements, which are most efficiently driven out of galaxies by prompt SNe-II (see \citealt{Yates+13}). Heavier $\alpha$ elements, such as silicon, have more similar CGM abundances in our default and modified models, with a difference in [Si/H] of only $\sim{}0.06$ dex at the same redshift and mass. Therefore, observational analyses that focus only on heavier alpha elements and assume solar relative abundances could under-estimate the overall \Zht{} in the CGM (see also \citealt{Wotta+19}). Nonetheless, this comparison with COS-Halos data suggests that the value of $f\sub{SNII,hot} = 0.9$ used in the modified model is close to its maximum permissible value.

\subsubsection{IGrM and ICM metallicities at low redshift}\label{sec:ZICM}
In higher-temperature systems, such as galaxy groups and clusters,\footnote{defined as systems with $13\lesssim{}\tn{log}(M\sub{vir}/\Msun)\lesssim{}14$, and $\tn{log}(M\sub{vir}/\Msun)\gtrsim{}14$, respectively.} emission lines from key metal ions such as iron become prominent in X-ray spectra, allowing for more precise measurements of the metal abundance to be made.

\begin{figure}
\centering
 \includegraphics[angle=0,width=0.99\linewidth]{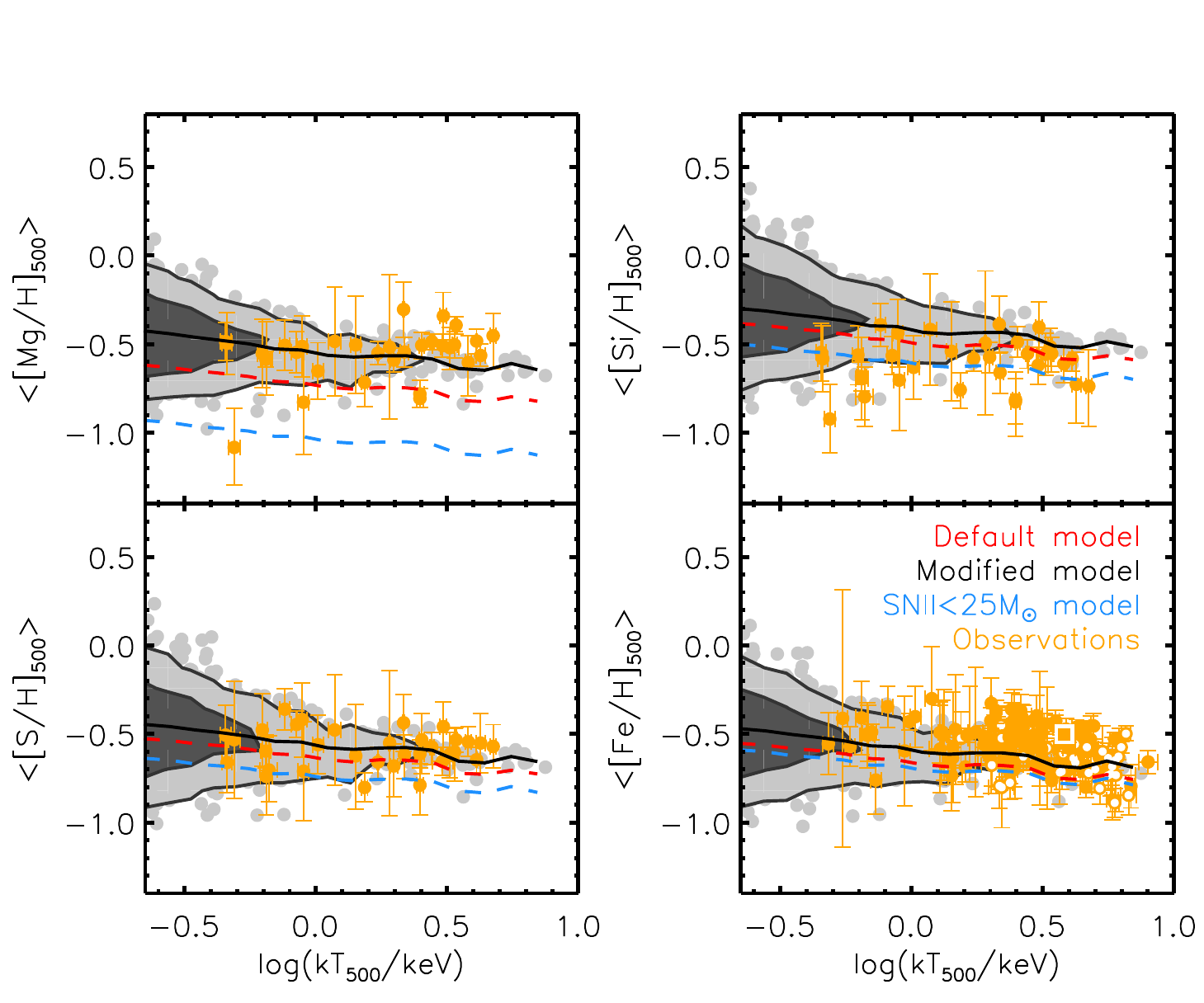}
 \caption{The relation between temperature at $r_{500}$ and the mean abundance within $r_{500}$ of [Mg/H], [Si/H], [S/H], and [Fe/H] in the ICM (\ie{}the T\ZICMt{}R). Grey contours denote systems from the modified \lgaltt{} model, with the mean relation given by the black solid lines. The mean relation for the default model (red dashed lines) and MaxGCEMass25 model (blue dashed lines) are also shown. Observational data include: homogenised [Fe/H] measurements for clusters from \citet{Yates+17} (open orange circles), abundances for all four chemical elements from the CHEERS sample \citep{Mernier+18a,Mernier+18b} (filled orange circles), and the [Fe/H] measured by \textit{Hitomi} for the Perseus cluster \citep{Simionescu+19} (white-edged square).}
 \label{fig:TZRs_ICM_incSNIIMAX25}
\end{figure}

Fig. \ref{fig:TZRs_ICM_incSNIIMAX25} shows the mass-weighted chemical abundances of Mg, Si, S, and Fe in the intra-group medium (IGrM) and ICM, as a function of the temperature at $r_{500}$ (\ie{}the T\ZICMt{}R). The mean relations for the default model (red) and modified model (black) are shown, along with the mean relation for the MaxGCEMass25 model (blue) discussed in Section \ref{sec:Low-SNII-max model}. A compilation of observational data from \citet{Yates+17} (open orange circles) and \citet{Mernier+18a,Mernier+18b} (filled orange circles) are also shown, along with the recent measurement of the iron abundance in the Perseus cluster from \textit{Hitomi} by \citet{Simionescu+19} (white-edged square). Here, radial gradients in temperature, gas density, and metallicity are accounted for, with both model and observational data being homogenised and re-scaled to $r_{500}$ following the techniques presented in \citet{Yates+17} and updated in section 4.1.3 of \citet{Henriques+20}.

Fig. \ref{fig:TZRs_ICM_incSNIIMAX25} shows that the modified model matches well to the chemical abundances observed in the IGrM and ICM for all four elements considered here. The higher abundances seen compared to the default model are again due to the increased direct CGM enrichment by SNe-II and SNe-Ia. We note that most significance should be given to the [Si/H], [S/H], and [Fe/H] abundances here, as Mg measurements from X-ray spectra can be significantly affected by both background and instrumental effects beyond the core region (F. Mernier, priv. comm.). In order to minimise such biases, we have adopted the shape parameters ($x\sub{c}$ and $\alpha$) obtained from the beta-profile fit to the [Fe/H] profile when re-scaling [Mg/H] here, while still allowing the normalisation to remain free (see \citealt{Yates+17}, section 2.4). Nonetheless, we advise caution when comparing Mg abundances between models and observations.


From Fig. \ref{fig:TZRs_ICM_incSNIIMAX25} we conclude that the modified \lgaltt{} model does even better than the default model in reproducing the typical metal content seen in the hot gas surrounding groups and clusters at $z=0$. This is a further indication that the increased direct CGM enrichment implemented in the modified model works well for metallicities both inside and outside galaxies.

\subsection{Evolution of global metallicities}\label{sec:Z_evo}

\begin{figure}
\centering
 \includegraphics[angle=0,width=0.8\linewidth]{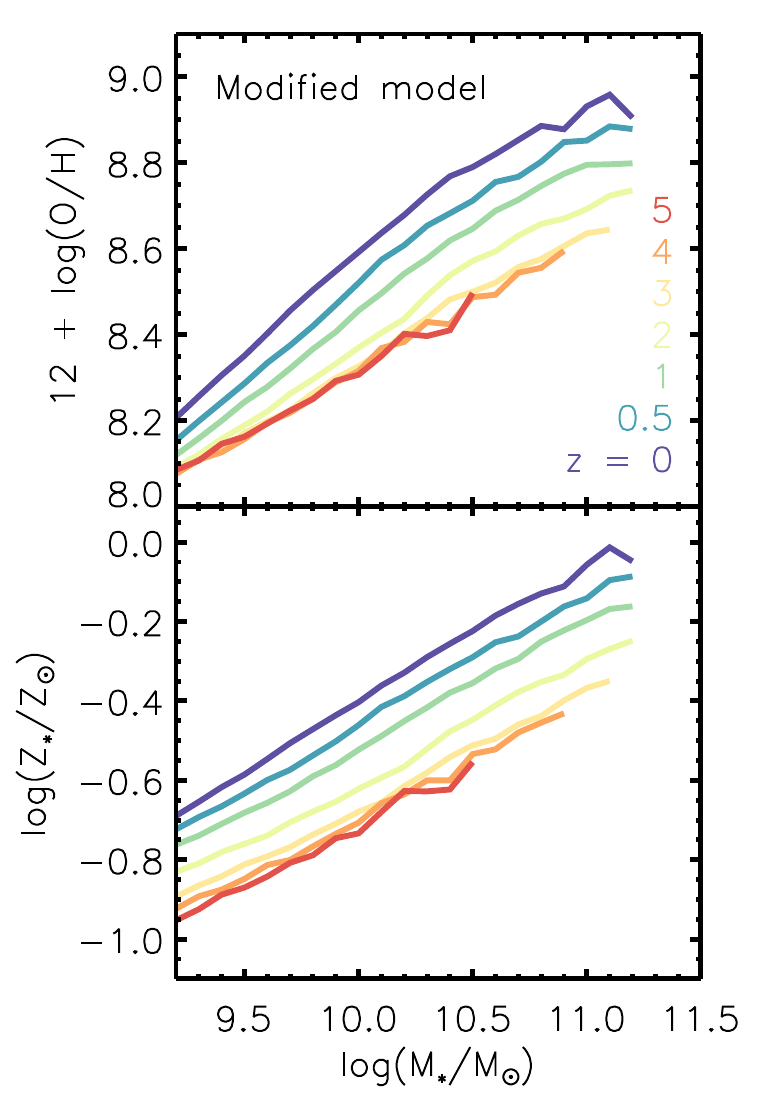}
 \caption{Evolution of the M\Zgt{}R (top panel) and M\Zst{}R (bottom panel) from redshift 5 to 0 in the modified \lgaltt{} model for model galaxies selected at each redshift to be star-forming and disc-dominant.}
 \label{fig:MZgR_and_MZsR_evo}
\end{figure}

The top panel of Fig. \ref{fig:MZgR_and_MZsR_evo} shows the evolution of the M\Zgt{}R for the modified model from $z=5$ to the present day. This model exhibits a clear increase in \Zgt{} at fixed stellar mass over cosmic time, caused by the efficient removal of newly-formed metals out of galaxies. In contrast, our default model exhibits a negligible evolution in \Zgt{} at high mass, and even a mildly \textit{inverted} evolution at low mass, due to an over-enrichment of the ISM at early times. Such issues have been common to older galaxy evolution models (see \eg{}\citealt{Yates+12,Fu+12,Guo+16,Knebe+18}), particularly those which do not calibrate to higher-redshift observational data (D. Stoppacher, priv. comm.).


For the modified model, we find an increase in mean \Zgt{} between $z=3$ and 0 at \logM{} $= 10.0$ of $\sim{}0.3$ dex. This is in good agreement with other semi-analytic and hydrodynamical models, which find a typical evolution in \Zgt{} of 0.25 - 0.45 dex for the same redshift range and stellar mass (\eg{}\textsc{Illustris}, \citealt{Torrey+14}; \textsc{Santa Cruz}, \citealt{Somerville+15}; \textsc{Gaea}, \citealt{Hirschmann+16}; \textsc{Fire}, \citealt{Ma+16}; \textsc{Mufasa}, \citealt{Dave+17}; \textsc{Eagle}, \citealt{DeRossi+17}; \textsc{Sag}, \citealt{Collacchioni+18}; \textsc{Simba}, \citealt{Dave+19}; \textsc{Illustris-TNG}, \citealt{Torrey+19}). However, some of these theoretical studies rely on explicit redshift dependencies to drive an increase in the normalisation of the M\Zgt{}R over cosmic time. The modified \lgaltt{} model, on the other hand, is able to reproduce a clear evolution in \Zgt{} at fixed mass by using a standard $V\sub{max}$-dependent reheating efficiency and a large, fixed direct CGM enrichment efficiency.

\begin{figure}
\centering
 \includegraphics[angle=0,width=0.99\linewidth]{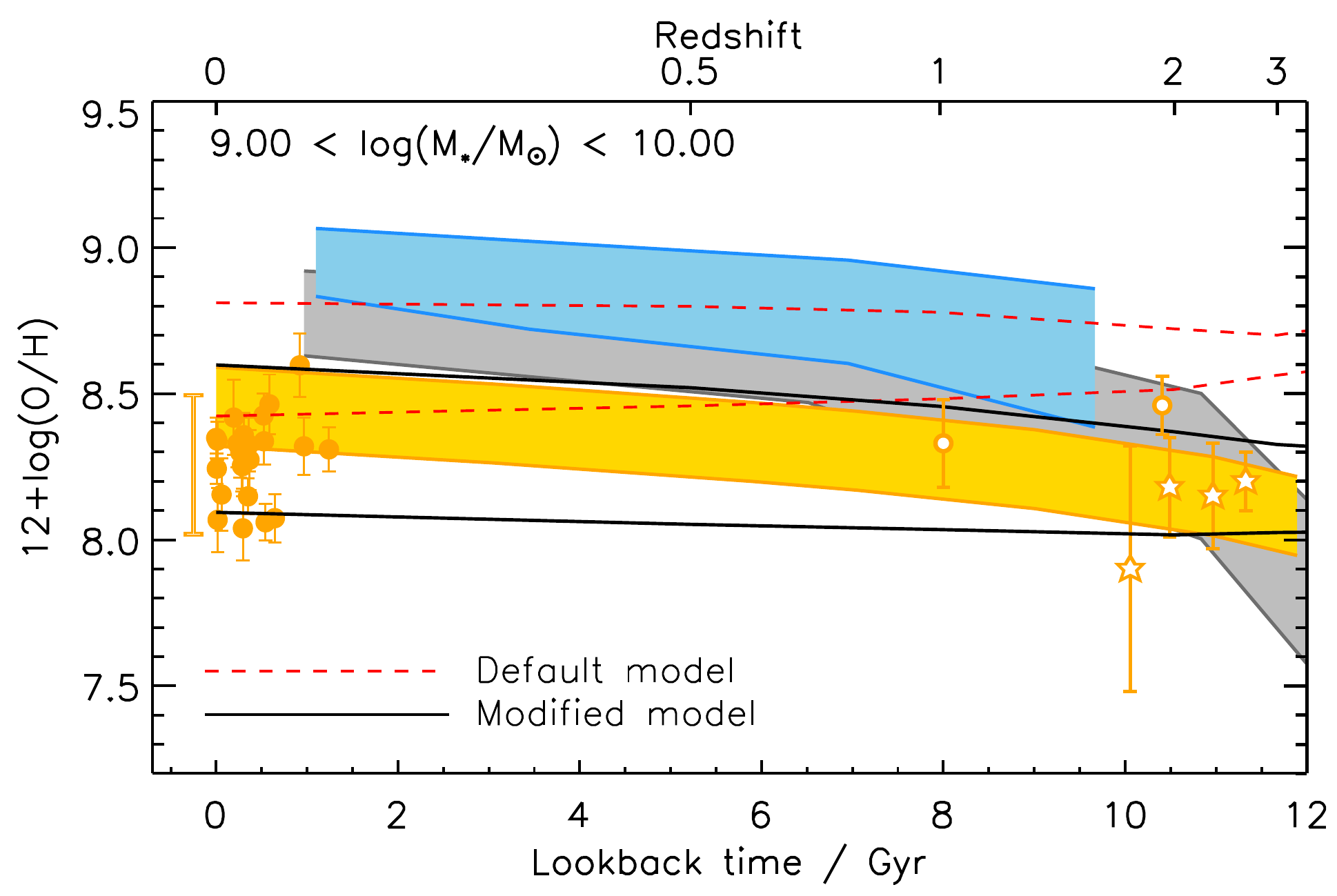}
 \caption{The evolution of ISM metallicity (\Zgt{}) in star-forming galaxies as a function of lookback time, for galaxies in the mass range $9.0 \leq \logMm{} < 10.0$. Black solid lines bracket the range of mean \Zgt{} for this mass range from the modified \lgaltt{} model. Red dashed lines indicate the same for the default model. Shaded regions denote the evolution in \Zgt{} inferred from the M\Zgt{}R studies of \citeauthor{Maiolino+08} (2008, grey), \citeauthor{Zahid+14} (2014, blue), and \citeauthor{Hunt+16} (2016, yellow). The orange/white bar (offset to slightly negative redshift for clarity) represents the range of \Zgt{} found for galaxies in this mass range from our MaNGA sample. Filled orange circles represent individual systems with \Te{}-based \Zgt{} estimates from \citet{Yates+20}. Open orange circles represent DLA systems (quasar IDs: Q0302-223 and Q2206-199) with absorption-line-based \Zgt{} from \citet{DeCia+18}. Open orange stars represent GRB host galaxies (GRB050820A, GRB081008, GRB120119A, and GRB121024A) with absorption-line-based \Zgt{} from \citep{Wiseman+17a} and stellar masses from \citet{Kruehler+15,Perley+16b}.}
 \label{fig:OH_evo}
\end{figure}

Fig. \ref{fig:OH_evo} shows the evolution of \Zgt{} for the modified model in comparison to various observational studies at different redshifts. This comparison is made for a fixed stellar mass bin of $9.0 \leq \logten{}(M_{*}/\Msun) < 10.0$, to mitigate the strong dependence on stellar mass that both metallicity and detectability can have. The modified \lgaltt{} model is shown by black solid lines, which bracket the range of \textit{mean} \Zgt{} present in galaxies within our chosen mass range. The same range for the default model is shown by red dashed lines.

The equivalent range of \Zgt{} from three different observational studies utilising SL \Zgt{} diagnostics are also shown: \citet{Maiolino+08} (grey shaded region, using a combination of KD02-combi, M08-O3, and M08-N2), \citet{Zahid+14} (blue shaded region, using KK04-R23 and PP04-N2), and \citet{Hunt+16} (yellow shaded region, using various diagnostics re-scaled to PP04-N2). Our MaNGA sample at $z\sim{}0.037$ is also shown by the orange bar, offset to negative redshift in Fig. \ref{fig:OH_evo} for clarity. Finally, a small collection of galaxies within the chosen mass range with directly-measured ISM metallicities are also shown: galaxies with electron-temperature based \Zgt{} from \citet{Yates+20} (filled orange circles), two damped Lyman alpha (DLA) systems from \citeauthor{DeCia+18} (2018, kindly provided by C. P\'{e}roux, priv. comm.) (open orange circles, one a lower limit due to its large impact parameter), and four gamma-ray burst (GRB) host galaxies with \Zgt{} measurements from multiple metal absorption lines \citep{Wiseman+17a} (open orange stars). These systems all have robust measurements for their host stellar mass and uncertainties in their \Zgt{} estimates of $<0.45$ dex.

The modified model is in relatively good agreement with the ensemble of direct metallicity measurements compiled here, although the number of such observed systems is currently quite small. The modified model also agrees well with the observed evolution reported by \citet{Hunt+16}. The PP04-N2 diagnostic used in that study agrees reasonably well with D16-NS2 at $z=0$ for galaxies with \logM{} $\sim{}10.0$, although less so at higher and lower masses.

The other two observational studies shown in Fig. \ref{fig:OH_evo} suggest much higher \Zgt{} at low redshift, but similar \Zgt{} at high redshift, and therefore a greater overall evolution over cosmic time. Interestingly, the majority of previous galaxy evolution models have compared and/or calibrated their M\Zgt{}Rs at $z=0$ to SL diagnostic measurements similar to those used by \citet{Maiolino+08} and \citet{Zahid+14}. This approach has lead to the conclusion that the \Zgt{} evolution of $\sim{}0.3$ dex seen in models could be insufficient. However, an important result of the work presented here is that the rate of \Zgt{} evolution seen in the majority of cosmological galaxy evolution models, including our modified model, is in fact in good agreement with the latest observational data. Most recently, \citet{Sanders+20} have measured the M\Zgt{}R evolution from $z\sim{}3.3$ to 0 using the MOSDEF survey, and have found an average increase in \Zgt{} of 0.35 dex at $\logMm{} = 10.0$, also in good agreement with our modified \lgaltt{} model.

The bottom panel of Fig. \ref{fig:MZgR_and_MZsR_evo} shows the evolution of the M\Zst{}R from $z=5$ to 0 for the modified model (with the mean mass-weighted \Zst{} from all radial rings considered here). Similarly to the M\Zgt{}R, we see a consistent evolution at fixed mass over cosmic time. There is an increase in \Zst{} of $\Delta{}(\Zs{})\sim{}0.24$ dex from $z\sim{}2$ to 0 in the modified model, which is in good agreement with the combined evolution observed by \citet{,Ferreras+19} and \citet{Gallazzi+14}, who found $\Delta{}(\Zs{})\sim{}0.07\error{0.07}$ dex from $z\sim{}2$ to 0.5 and $\Delta{}(\Zs{})\sim{}0.12\error{0.05}$ dex from $z\sim{}0.5$ to 0.1, respectively, for galaxies with $\logMm{}\sim{}11.0$.

We also find a much reduced evolution in the M\Zst{}R for ETGs compared to star-forming galaxies in \lgaltt{}, in agreement with observations (\eg{}\citealt{Kriek+19,Estrada-Carpenter+19,Leethochawalit+19,Lonoce+20}). At $\logMm = 11.0$, \Zst{} increases by only 0.04 dex from $z=2$ to 0 for ETGs in the modified model. This is due to the relative lack of metal production in such quiescent systems at late times.


Regarding CGM metallicities at higher redshift, the general consensus from X-ray studies of the IGrM and ICM is that a significant fraction of the present-day iron abundance was already in place by $z\sim{}1-2$ (\eg{}\citealt{Balestra+07,Anderson+09,McDonald+16}). For example, \citet{McDonald+16} conclude that the metallicity (\ie{}the metal to hydrogen ratio) in the ICM at $z=1$ was already at least 60 per cent of that seen in clusters today. This scenario is in line with the early build-up of metals also seen in the stellar populations of ETGs discussed above.

In \lgaltt{}, both the default and modified models broadly reproduce this observed trend for the ICM. In the default model, the mean `iron abundance fraction' within $r_{200}$ (\ie{}$\tn{[Fe/H]}\sub{200,z}/\tn{[Fe/H]}\sub{200,z=0}$) for clusters with log$(\tn{k}T\sub{500,z=0}/\tn{keV}) > 0.1$ is already 0.80 at $z=1$ and increases to 0.88 by $z=0.5$. This is similar to the result found by \citet{Yates+17}, who studied an earlier version of \lgal{} which assumed $f\sub{hot}$ values of 0.0.

The modified \lgaltt{} model shows an even smaller evolution at late times, with $\tn{[Fe/H]}\sub{200,z}/\tn{[Fe/H]}\sub{200,z=0} = 0.84$ at $z=1$ and 0.89 at $z=0.5$. This puts the modified model in somewhat better agreement with the observed rate of the ICM iron abundance evolution, notwithstanding the very large scatter in observed $\tn{[Fe/H]}\sub{200}$ measurements (see \citealt{Yates+17}, section 6.5).

\begin{figure*}
\centering
 \includegraphics[angle=0,width=0.99\linewidth]{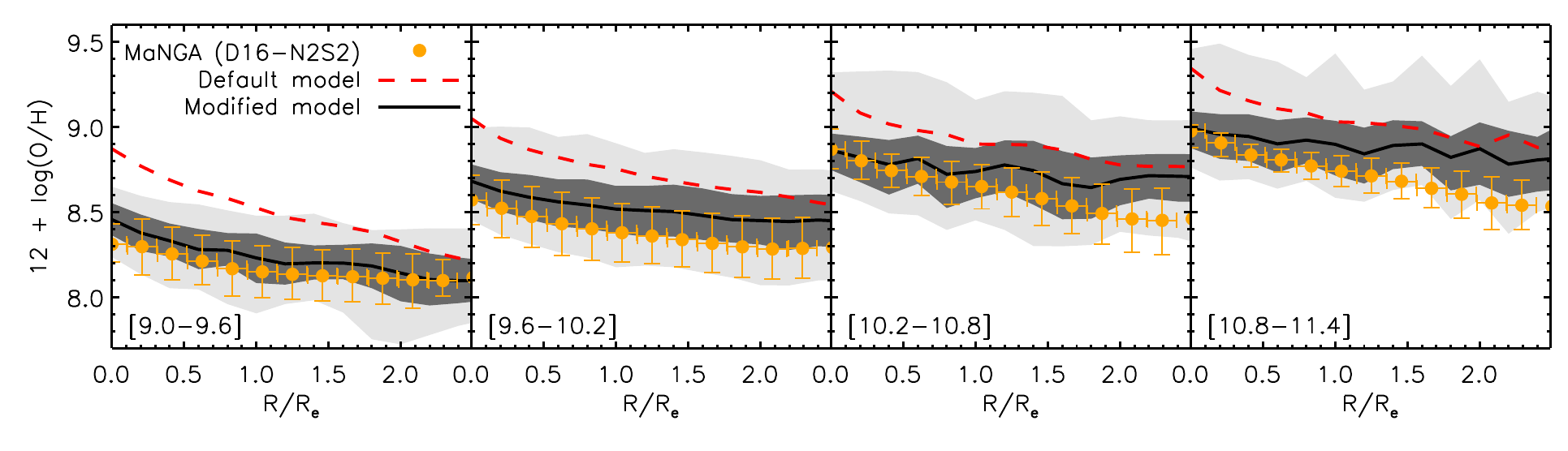}\\
 \includegraphics[angle=0,width=0.765\linewidth]{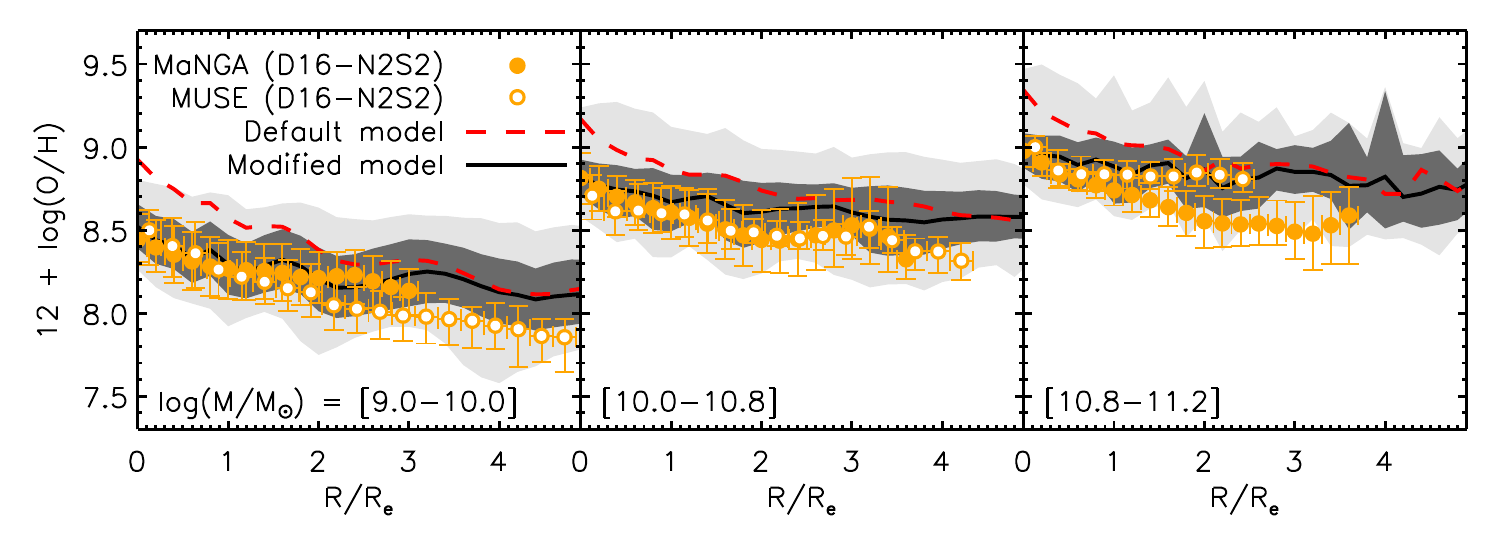}
 \caption{\textit{Top panels:} Radial \Zgt{} profiles out to 2.5\Reff{} for galaxies in four mass bins. Black solid lines represent the modified model (with dark and light grey contours representing the 1$\sigma$ and 2$\sigma$ spread, respectively), red dashed lines represent the default model, and filled orange circles represent our MaNGA sample (with vertical bars representing the 1$\sigma$ spread in \Zgt{}). \textit{Bottom panels:} The same as the top panels, but with profiles shown out to out to 5\Reff{}. Data is re-binned here into only three mass bins, in order to compare with the data from \citet{Erroz-Ferrer+19} for the MUSE/MAD sample (open orange circles).}
 \label{fig:Zg_profiles_z0}
\end{figure*}


\begin{figure*}
\centering
 \includegraphics[angle=0,width=0.99\linewidth]{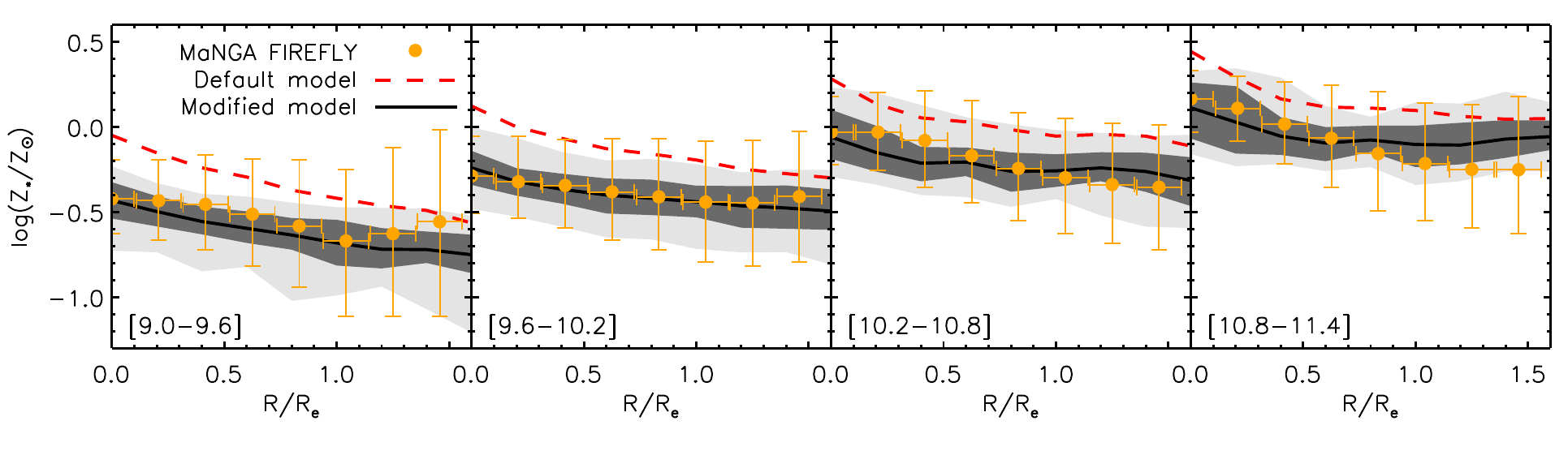}
 \caption{Radial \Zst{} profiles out to 1.6\Reff{} for galaxies in four mass bins. Lines, contours, and colours are as in Fig. \ref{fig:Zg_profiles_z0}. Radial profiles for the MaNGA survey are constructed using the \Zst{} measurements made using FIREFLY \citep{Goddard+17}.}
 \label{fig:Zstar_profiles_z0}
\end{figure*}

In conclusion, we find that the modified \lgaltt{} model is in better agreement with a range of higher-redshift global metallicity observations than earlier versions of the model. These improvements are primarily due to the implementation of increased direct enrichment of the CGM by SNe, which decreases the amount of metal mixing in the ISM and therefore lowers the gas and stellar metallicities inside galaxies, while increasing the metal content present in the hot gas surrounding galaxies, groups, and clusters.

\subsection{Metallicity profiles at redshift zero}\label{sec:Profiles at z=0}

In the upper panels of Fig. \ref{fig:Zg_profiles_z0}, we show the stacked radial \Zgt{} profiles out to 2.5 \Reff{} for star-forming disc galaxies, split into four mass bins. For each mass bin, average profiles are obtained by taking the mean metallicity in annuli of 0.2 $R/\Reffm{}$ width. Solid black lines and dashed red lines represent the mean profile from the modified and default \lgaltt{} models, respectively. Filled orange points represent the mean profile from our MaNGA sample, with \Zgt{} calculated using the D16-N2S2 diagnostic and the vertical bars representing the 1$\sigma$ spread in \Zgt{} measurements in each radial bin.

There is promising agreement between the radial profiles in the modified model and the MaNGA observations. This is partly by construction, as we have used the \textit{normalisation} of the observed low-redshift metallicity profiles as a guide when setting the values of the modified GCE parameters (see Section \ref{sec:GCE parameters}). This shows that there is a clear improvement in the match between observed profiles and \lgaltt{} when a significant amount of direct CGM enrichment is allowed.

In the highest-mass bin, the mean profile for the modified model is flatter than seen in our MaNGA sample. However, other observational data suggests that such flatter profiles are indeed present in massive disc galaxies. This is shown in the lower panels of Fig. \ref{fig:Zg_profiles_z0}, where \Zgt{} profiles from the MUSE Atlas of Discs (MAD) sample are also shown (\citealt{Erroz-Ferrer+19}, open orange circles). That study utilised MUSE observations of 38 spiral galaxies at $z<0.013$ which lie on the star-forming main sequence and have inclinations of $<70^{\circ}$, to measure \Zgt{} profiles using the same D16-N2S2 diagnostic used here for our MaNGA sample. The MAD sample exhibits flatter radial profiles beyond $\sim{}0.3R\sub{e}$ in massive galaxies, while corroborating the steeper profiles seen in our MaNGA sample at intermediate masses.

We note here that the MaNGA and MUSE IFUs have different spatial resolutions. The lower resolution of MaNGA (equating to $\sim{}1$-2 kpc for our MaNGA sample, compared to an average of $\sim{}100$ pc for the MAD sample) could become problematic when measuring \Zgt{} at large galactocentric radii, as S/N tends to be lower and contamination from diffuse ionised gas (DIG) can become more significant (see also \citealt{Poetrodjojo+19}). However, lower resolution should artificially \textit{flatten} radial profiles, rather than steepen them (see \eg{}\citealt{Acharyya+20}). In the case of the MUSE data, \citet{Erroz-Ferrer+19} were able to disentangle the contributions from \HII{} regions and DIG in their sample, due to the higher resolution. Therefore, we tentatively conclude that the radial profiles provided by MUSE are more representative of the true distribution of metals in the ISM of massive disc galaxies, and find it promising that the modified \lgaltt{} model reproduces these profiles best.

Fig. \ref{fig:Zstar_profiles_z0} shows the radial profiles for stellar metallicity in disc galaxies, again split into four mass bins. We find good agreement between the modified \lgaltt{} model and the profiles from our MaNGA sample, particularly within $\sim{}1 \Reffm$, although we note the large spread in the observed \Zst{} at fixed radius. The modified model (black solid line) is again a better fit to the data than the default model (red dashed line). However, as with the \Zgt{} profiles, the modified model exhibits flatter \Zst{} profiles at large radii in massive galaxies than is seen in our MaNGA sample. It is again unlikely that lower spatial resolution is causing artifically steepened \Zst{} profiles in our MaNGA data. Therefore, this discrepancy between model and observations could instead indicate a shortcoming of our modified model in reproducing the metal content in discs at large radii (see Section \ref{sec:Low-SNII-max model}).

We note here that the stellar bulge component is not included when plotting the radial \Zst{} profiles for our model galaxies in Fig. \ref{fig:Zstar_profiles_z0}. This is because bulges are not yet resolved into self-consistent radial rings in \lgaltt{}, so their spatial metallicity distribution is unknown. This issue is somewhat mitigated by the fact that we only consider model systems with relatively minor bulges, requiring a bulge-to-total mass ratio of $< 0.3$ (see Section \ref{sec:Model sample selection}). Likewise, we also expect the contribution from bulges in our comparison MaNGA sample to be minimal, as these systems have been specifically selected to have high sSFR, an inclination of $<60^{\circ}$, and a clear spiral morphology from visual inspection (see Section \ref{sec:Sample selection}).


\subsection{Are metallicity profiles mass dependent?}\label{sec:Mass dependent Zg profiles}
There is currently no clear consensus in the observational literature as to whether \Zgt{} profiles are dependent on stellar mass in nearby galaxies. Using a variety of different IFUs and \Zgt{} diagnostics, some observational studies have suggested steeper slopes in more massive galaxies (\eg{}\citealt{Belfiore+17,Poetrodjojo+18}),
others flatter slopes in more massive galaxies (\eg{}\citealt{Kaplan+16,Erroz-Ferrer+19}), and others no clear mass dependence at all (\eg{}\citealt{Sanchez+14,Lian+18b}). This provides a good opportunity for galaxy evolution models such as \lgaltt{} to investigate metallicity profiles, as they work from a set of fundamental theoretical assumptions to provide a physical explanation for a given trend that is independent of observational biases.

\begin{figure}
\centering
 \includegraphics[angle=0,width=0.99\linewidth]{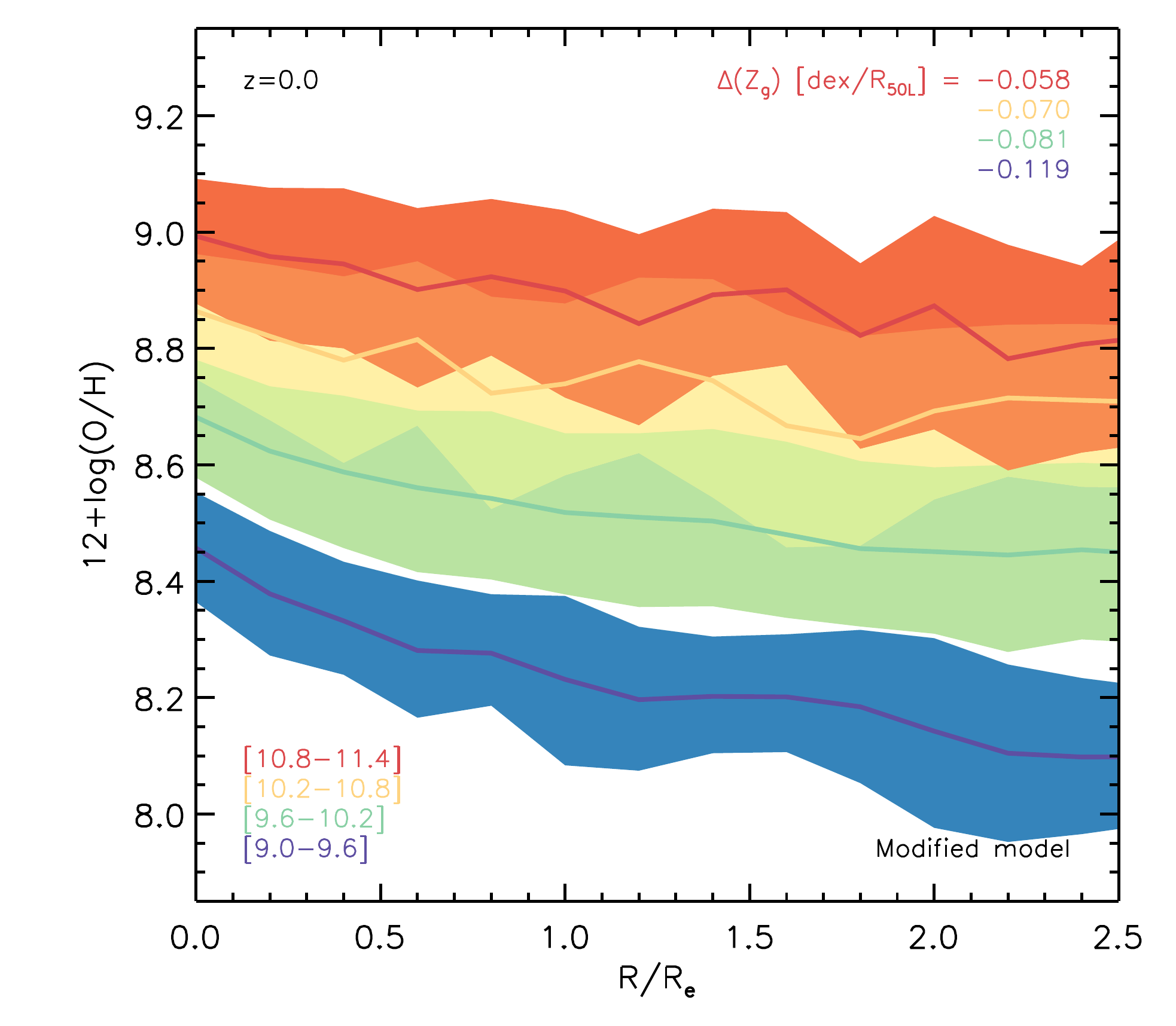}
 \caption{Mean radial \Zgt{} profiles for star-forming, disc galaxies at $z=0$ from our modified model, split into four mass bins. The 1$\sigma$ spreads are illustrated by shaded regions. The mean slope in dex/\Reff{} from linear fits to the \Zgt{} profile between 0.5 and 2$R\sub{e}$ for each galaxy are reported for each mass bin in the top-right corner.}
 \label{fig:Mass-dep profiles}
\end{figure}

Fig. \ref{fig:Mass-dep profiles} shows the binned \Zgt{} profiles for $z=0$ disc galaxies in the modified \lgaltt{} model with all four mass bins plotted together. We have also calculated a linear fit to every model galaxy's \Zgt{} profile between 0.5 and 2\Reff{}, and report the \textit{mean} of the slopes in dex/\Reff{} for each mass bin in the top right corner. There is a clear, yet rather weak, dependence of the slope, $\Delta(\Zg{})$, on stellar mass in the modified model, such that lower-mass galaxies have steeper $\Delta(\Zg{})$. A similar mass dependence is also seen for the model \Zst{} slopes.

\begin{figure}
\centering
 \includegraphics[angle=0,width=0.95\linewidth]{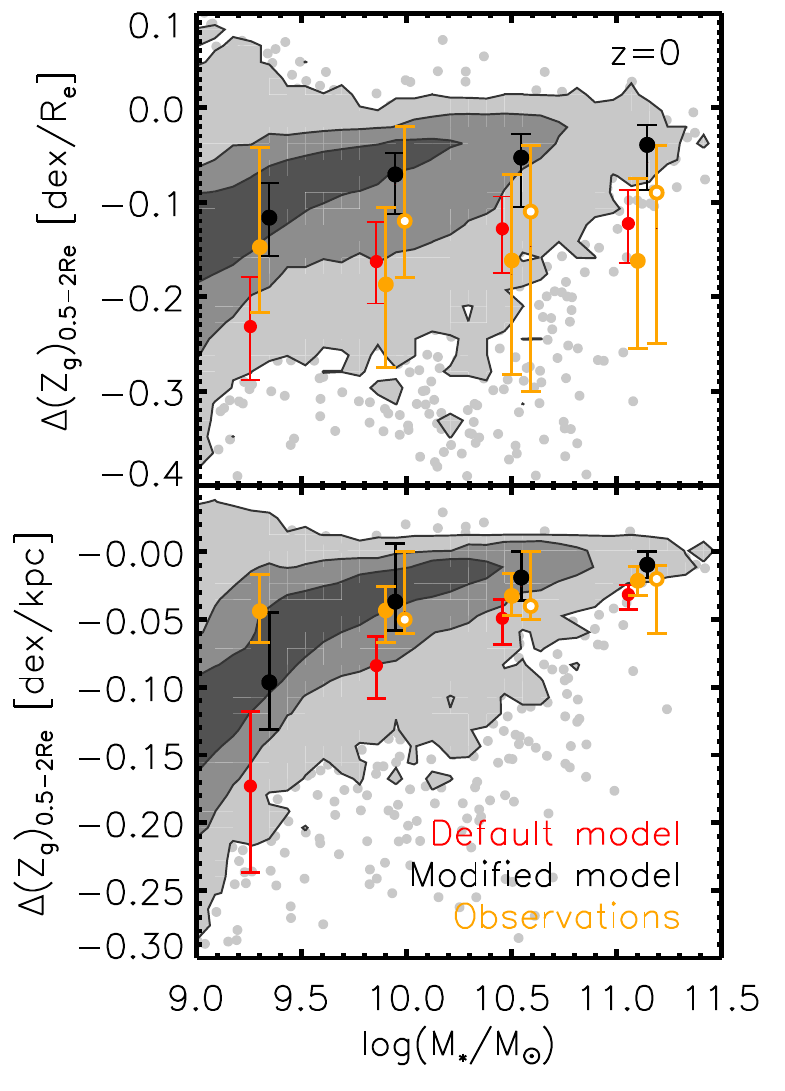} 
 \caption{\textit{Top panel:} \Zgt{} profile slope, $\Delta(\Zg{})$, in units of dex/\Reff{} as a function of stellar mass for star-forming disc galaxies at $z=0$. Grey contours and points represent the whole sample from the modified model. Red and black points represent the median $\Delta(\Zg{})$ fit between 0.5 and 2\Reff{} for the default and modified \lgaltt{} model, respectively. Filled orange circles represent the same median $\Delta(\Zg{})$ from our MaNGA sample. Open orange circles represent the median $\Delta(\Zg{})$ from the MUSE/MAD sample \citep{Erroz-Ferrer+19}. In all cases, vertical bars represent the 16th-84th percentile range in $\Delta(\Zg{})$ for each mass bin. \textit{Bottom panel:} Same as top panel, but with slopes measured in dex/kpc.}
 \label{fig:M-ZgGrad_z0}
\end{figure}

This trend is more clearly represented in Fig. \ref{fig:M-ZgGrad_z0}, where $\Delta(\Zg{})$ is shown as a function of stellar mass. Slopes are measured in dex/\Reff{} in the top panel, and dex/kpc in the bottom panel. Filled circles represent the \textit{median} $\Delta(\Zg{})$ in four mass bins from the modified model (black), default model (red), our MaNGA sample (filled orange), and the MAD sample (open orange). For the model and MaNGA samples, the median values have been calculated following the method of \citet{Belfiore+17}, by azimuthally averaging the \Zgt{} in radial bins for each galaxy, then taking its slope from an unweighted linear fit between 0.5 and 2\Reff{} (considering only galaxies with `clean' \Zgt{} measurements in four or more radial bins in this range), and then taking the median of these slopes in each mass bin. For the MAD sample, we take the individual slope measurements provided by \citealt{Erroz-Ferrer+19} in their table 1. In this case, only galaxies with \Zgt{} measurements for \HII{} regions in two or more 0.3-dex-wide radial bins between 0.5 and 2\Reff{} were considered.

When comparing the default and modified models to each other, we see that \Zgt{} gradients are overall flatter (although still predominantly negative) in the modified model, especially at lower mass. This is caused by the enhanced direct CGM enrichment, as described in detail in the next section.

Regarding the observations, we find that the median $\Delta(\Zg{})$ for low-mass galaxies in our MaNGA sample is particularly sensitive to (a) the SL \Zgt{} diagnostic chosen, and (b) whether or not a spiral morphology is required. For example, when measured in dex/\Reff{}, the median slope for the lowest-mass bin flattens from $-0.148$ when using the D16-N2S2 diagnostic to $-0.068$ when using the M08-R23 diagnostic. Similarly, this median value decreases from $-0.148$ to $-0.098$ when not selecting on morphology. This latter change is likely due to a significant number of low-mass irregular galaxies entering the sample, which do not contain well-structured discs from which meaningful radial \Zgt{} profiles can be obtained. We emphasise here that the clear majority of low-mass galaxies in MaNGA appear to be of non-spiral morphology. For example, the number of MaNGA galaxies in our lowest two mass bins increases by a factor of 5.0 and 2.7, respectively, when allowing all morphologies into the sample. This increase is reduced to factors of 1.4 and 1.5, respectively, for the highest two mass bins, reflecting the fact that massive star-forming galaxies more commonly have spiral morphologies. We also acknowledge that low-mass, disc-dominant galaxies which do not exhibit visible spiral structure may be excluded from our MaNGA sample.

The combination of using the M08-R23 \Zgt{} diagnostic and relaxing the spiral morphology requirement decreases the median $\Delta(\Zg{})$ for our lowest-mass bin to $-$0.058 dex/\Reff{}, with this value being just consistent with 0.0 within the 1$\sigma$ spread. This could therefore partly explain the near-flat median profiles reported for low-mass galaxies by \citet{Belfiore+17} when using the M08-R23 or PP04-O3N2 diagnostics and not explicitly selecting for spiral or disc-dominant systems. However, biases to the D16-N2S2 diagnostic due to residual dependences on SFR could also play a role (see Section \ref{sec:Metallicity diagnostics}).

The presence of galactic bars could also contribute to a flattening of \Zgt{} profiles in low-mass galaxies \citep{Zurita+21b}. Such dynamical components can funnel gas towards the centres of galaxies, redistributing their metals. This process can have a larger effect in systems with large bar-to-total stellar mass ratios. However, it is more likely that the residual flattening seen in the lowest-mass MaNGA bin in Fig. \ref{fig:M-ZgGrad_z0} is due to spatial resolution. In their hydrodynamical simulations of disc galaxies, \citet{Acharyya+20} found that measured \Zgt{} slopes can be artificially flattened by $>20$ per cent for data with the spatial resolution and average PSF of MaNGA, with this effect being stronger for systems which have steeper `true' profiles. This could play a role in explaining why our MaNGA profiles appear somewhat shallower than those of the MAD sample for the lowest-mass bin in Fig. \ref{fig:Zg_profiles_z0}.

The top panel of Fig. \ref{fig:M-ZgGrad_z0} indicates that \Zgt{} profile slopes in the modified model are flatter than in the observations when measured in dex/\Reff{}. However, we note that our MaNGA galaxies have systematically larger \Reff{} than those in both the MUSE sample and \lgaltt{} at fixed stellar mass (noting that the model reproduces the typical \Reff{} found in large populations of local, star-forming galaxies, see \citealt{Henriques+20}, section 4.2). The predominance of less compact stellar discs in our MaNGA sample is mainly due to the spiral morphology requirement, which preferentially removes low-mass, compact systems. Such an anti-correlation between disc size and $\Delta(\Zg{})$ in MaNGA is also found by \citet{Boardman+21}. Therefore, we also compare the median $\Delta(\Zg{})$ in units of dex/kpc in the bottom panel of Fig. \ref{fig:M-ZgGrad_z0}. We find an improved agreement between the modified model, the MUSE sample, and the MaNGA sample in this case, suggesting that part of the discrepancy seen when using dex/\Reff{} is indeed due to differences in \Reff{} at fixed mass among the samples.

When considering trends with stellar mass, we find that above \logM{} $\sim{}10.0$, all the observational and model samples studied here indicate a weak flattening of the slope with increasing stellar mass. The change in median $\Delta(\Zg{})$ from \logM{} $\sim{}9.9$ to 11.1 is quite similar between the modified model and observations, with 0.026 dex/\Reff{}/log($M_{*}$) for the modified model, compared to 0.025 dex/\Reff{}/log($M_{*}$) for the MAD sample and 0.021 dex/\Reff{}/log($M_{*}$) for our MaNGA sample. A similar agreement is seen when measuring slopes in dex/kpc.

Overall, our results are in best qualitative agreement with the findings from higher-resolution IFUs such as VIRUS-P and MUSE (\eg{}\citealt{Kaplan+16,Erroz-Ferrer+19}), where a weak positive correlation between $M_{*}$ and $\Delta{}(\Zg{})$ is seen. Also, both \citet{Pilyugin+19} and \citet{Bresolin19} report a weak flattening of \Zgt{} profiles with increasing stellar mass for a combination of long-slit, CALIFA, MaNGA, and MUSE spectra. Although, \citealt{Bresolin19} only find this trend when slopes are measured in dex/kpc (see also \citealt{Tissera+19}). Similarly, \citet{Zurita+21a,Zurita+21b} find a positive correlation between $\Delta{}(\Zg{})$ in dex/kpc and B-band luminosity for a large collection of \HII{} regions from 20 unbarred galaxies with $i < 70^{\circ}$. Their result holds when measuring \Zgt{} via either the \Te{} method or a range of SL \Zgt{} diagnostics.

At lower masses, the good agreement found between our results and those of the above studies could also be due to the selection of spiral galaxies in all cases. Studies that find gradients flattening at low masses (\eg{}\citealt{Belfiore+17}) tend to instead use the entire galaxy population without pre-selecting on morphology and, since most low-mass galaxies are irregular, their flat metallicity gradients dominate the statistics.

Given the weakness of the $M_{*}$ -- $\Delta(\Zg{})$ correlation found in \lgaltt{}, our results are also consistent with the largely mass-\textit{independent} slopes reported by \citet{Sanchez+14} and \citet{Sanchez-Menguiano+16} using CALIFA, \citet{Lian+18b} using MaNGA, and \citet{Carton+18} using spatially-binned MUSE spectra of galaxies at $0.1 \lesssim{} z \lesssim{} 0.8$.

In the following section, we discuss the evolutionary processes taking place in \lgaltt{} that lead to the metallicity profiles present at redshift 0.

\subsection{The evolution of metallicity profiles}\label{sec:Profiles evo}
In this section, we focus on results from \lgaltt{} run on the \textsc{Millennium-II} simulation, as this provides the higher resolution needed to study the low-mass progenitors of model galaxies back to high redshift.

\begin{figure}
\centering
 \includegraphics[angle=0,width=0.96\linewidth]{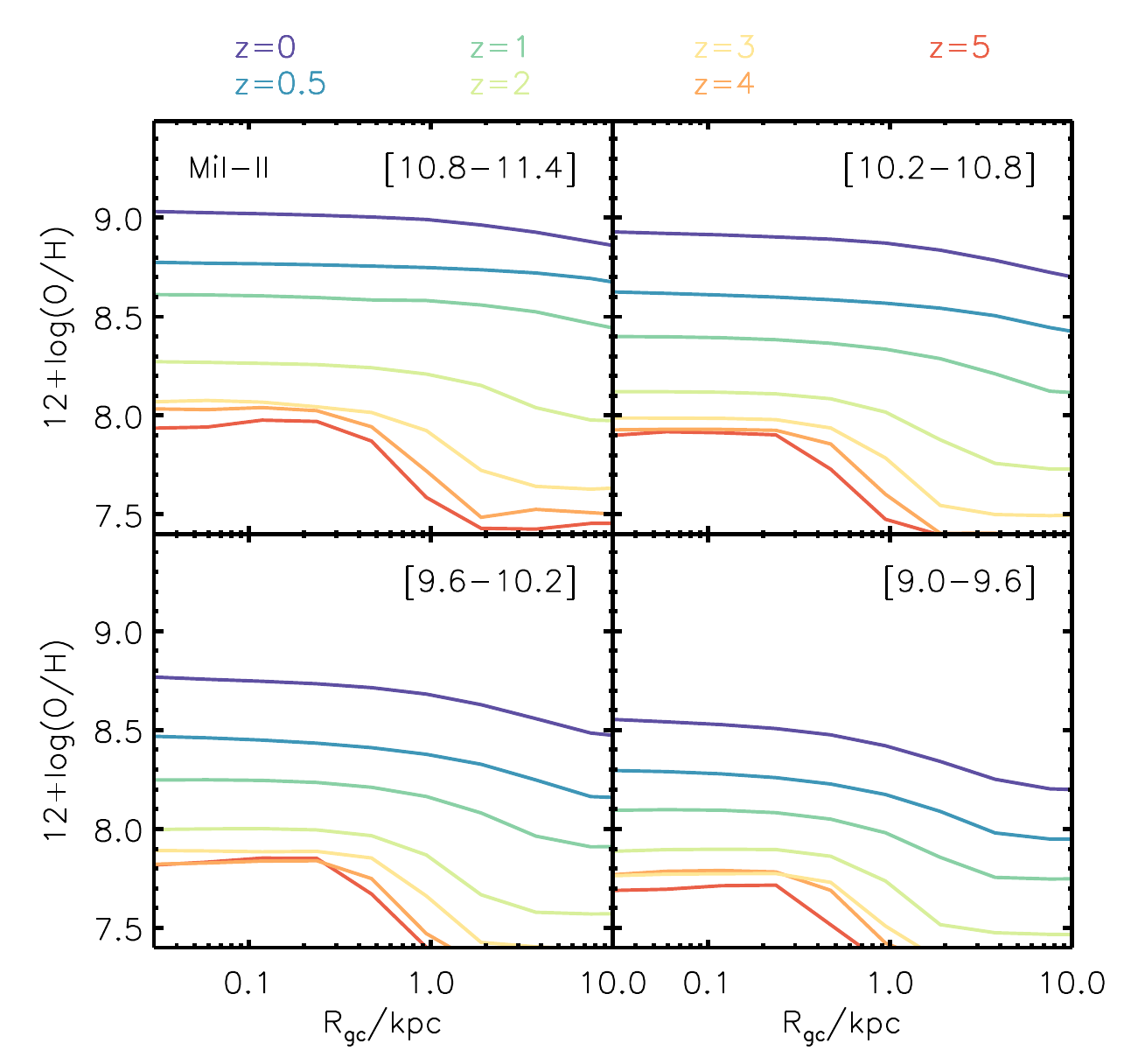}
 \caption{The evolution of radial \Zgt{} profiles (as a function of galactocentric radius) for star-forming, disc galaxies selected at $z=0$ in the modified model run on \textsc{Millennium-II}. Galaxies are split into four mass bins according to their stellar mass at $z=0$.}
 \label{fig:Zg_profiles_perRgc_evo}
\end{figure}

When studying an earlier version of \lgal{}, \citet{Fu+13} established that the steepness of \Zgt{} profiles depended on the merger history of the galaxy, as implied by its present-day bulge-to-total stellar mass ratio (B/T). When considering all model galaxies at $z=0$, they found that those with larger B/T ratios had flatter \Zgt{} profiles (when measured from the centre out to the far edge of the disc). In this present work, we instead focus only on disc-dominant (B/T $< 0.3$), star-forming galaxies in \lgaltt{}, and measure profiles within $0.5-2$ \Reff{}, in order to better compare to observations. For this sample, we do not find a strong correlation between B/T and $\Delta{}(\Zg{})$, indicating that it is mainly secular processes that determine the differences in \Zgt{} profiles seen in our model discs.

\begin{figure}
\centering
 \includegraphics[angle=0,width=0.99\linewidth]{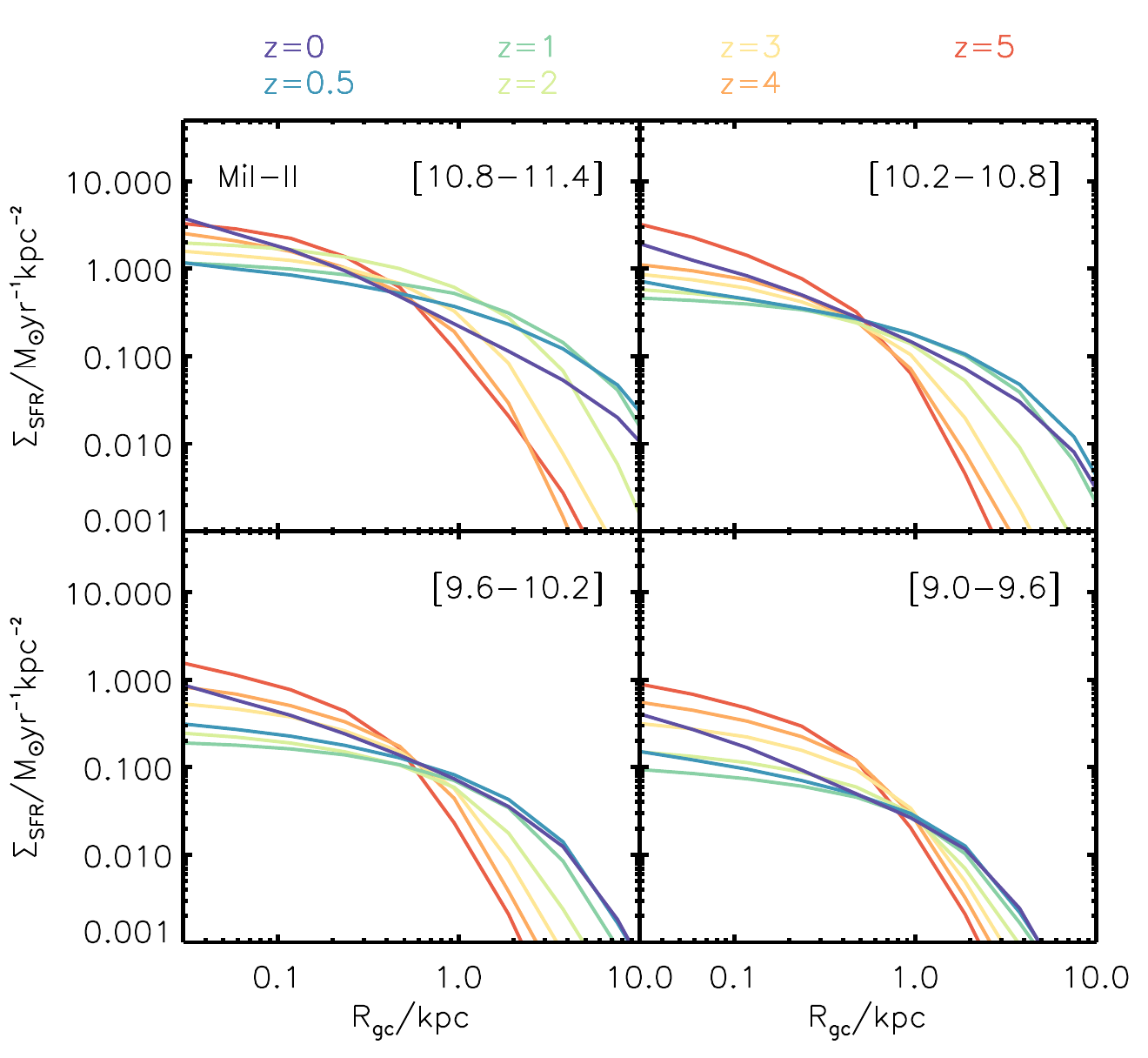}
 \caption{The evolution of radial $\Sigma\sub{\textsc{SFR}}$ profiles for the same model galaxies as shown in Fig. \ref{fig:Zg_profiles_perRgc_evo}.}
 \label{fig:SFR_profiles_perRgc_evo}
\end{figure}

\begin{figure}
\centering
 \includegraphics[angle=0,width=0.96\linewidth]{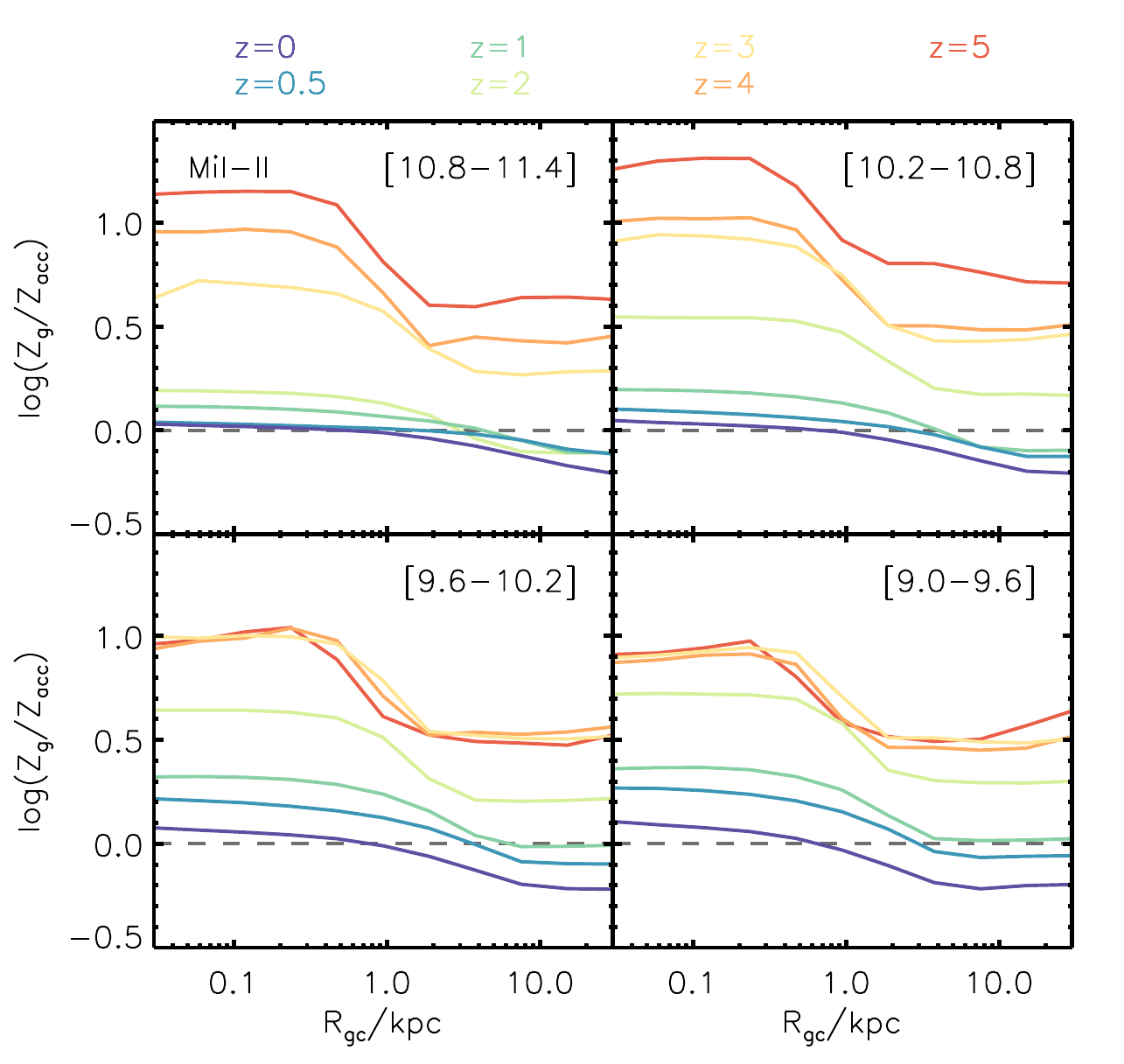}
 \caption{The evolution of the ratio of ISM oxygen abundance to accreted oxygen abundance ($\Zg/Z\sub{acc}$) for the same model galaxies as shown in Fig. \ref{fig:Zg_profiles_perRgc_evo}. Dashed grey lines indicate an equal oxygen abundance in the ISM and accreted gas. Values of log$(\Zg/Z\sub{acc})$ above 0 indicate dilution of the ISM, whereas values below 0 indicate enrichment of the ISM.}
 \label{fig:ZgZacc_profiles_perRgc_evo}
\end{figure}

There are three main secular processes driving the evolution of metallicity profiles in \lgaltt{}: \textit{inside-out growth} (\ie{}the spread of star formation out to larger radii over time), \textit{radial gas inflows}, and \textit{accretion of pre-enriched gas} from the CGM. The relative importance of these three processes determines the characteristic metallicity profiles seen at different masses and in different versions of the \lgaltt{} model.

Fig. \ref{fig:Zg_profiles_perRgc_evo} shows the evolution of mean \Zgt{} profiles as a function of galactocentric radius ($R\sub{gc}$) in the modified model, from redshift 5 to 0. In each mass bin, star-forming disc galaxies are selected at $z=0$, and then their main progenitors are traced back to higher redshifts in order to obtain the evolution of the \textit{same} systems over cosmic time. The evolution in \Zgt{} profiles seen in Fig. \ref{fig:Zg_profiles_perRgc_evo} can be best explained by considering the inner and outer disc separately. At low radii (below $\sim{}1$ kpc), on-going star formation (fuelled by radial inflows and cooling from the CGM) drives the increase in \Zgt{} seen over cosmic time. At large radii (above $\sim{}1$ kpc), inside-out growth and pre-enriched accretion drive the increase in \Zgt{}.

The effect of inside-out growth is best illustrated by Fig. \ref{fig:SFR_profiles_perRgc_evo}, which shows the evolution of SFR surface density profiles in our model disc galaxies. The $\Sigma\sub{\textsc{SFR}}$ increases beyond $\sim{}1$ kpc over cosmic time due to the formation of new H$_{2}$ at progressively larger radii as gas discs grow in size. This inside-out growth is significantly stronger in massive galaxies (top-left panel of Fig.~\ref{fig:SFR_profiles_perRgc_evo}) than their lower-mass counterparts (bottom-right panel of Fig.~\ref{fig:SFR_profiles_perRgc_evo}).

The effect of metal-rich accretion is best illustrated by Fig.~\ref{fig:ZgZacc_profiles_perRgc_evo}, which shows the ratio of the oxygen abundance in the ISM to that in the accreted gas, $\Zg{}/Z\sub{acc}$, as a function of radius. Due to the high direct CGM enrichment efficiency in the modified model (and the instantaneous mixing approximation assumed for the CGM in \lgaltt{}), gas accreted from the CGM can reach a higher metallicity than that in the low-density outskirts of galaxy discs. This leads to an enrichment at large radii, similar to the `galactic fountain' mechanism proposed by \citet{Fraternali+13}. In massive galaxies, such metal-rich accretion begins to occur from $z\sim{}2$ (top-left panel of Fig.~\ref{fig:ZgZacc_profiles_perRgc_evo}) and does so efficiently due to shorter cooling timescales. In low-mass galaxies, it does not begin until after $z\sim{}1$ (bottom-right panel of Fig.~\ref{fig:ZgZacc_profiles_perRgc_evo}) and does so at a lower rate due to longer cooling timescales.

We can therefore see that more significant in-situ star formation beyond $\sim{}1$ kpc, and more efficient metal-rich accretion at low redshift, drive a stronger increase in the outer \Zgt{} over time in massive disc galaxies than in their low-mass counterparts. This leads to the flatter \Zgt{} profiles seen in massive galaxies by redshift 0, and hence the weakly mass-dependent \Zgt{} slopes discussed in Section \ref{sec:Mass dependent Zg profiles}. A similar evolution is also seen for the outer discs of Milky-Way-type galaxies in the AURIGA simulation \citep{Grand+19}, and to a lesser extent in EAGLE \citep{Collacchioni+20}. Although, the latter study also reports steeper \Zgt{} profile slopes with increasing accretion rate for the inner disc.

In contrast to the modified model, star formation is the dominant driver of metal enrichment at all radii in the default model. 70 per cent of the oxygen synthesised and released by SNe is initially deposited into the ISM in the default model, allowing a lot of metals to immediately mix with the nearby star-forming gas. The gas accreted from the CGM is therefore on average more metal-poor than the ISM at all radii and redshifts, causing a net \textit{diluting} effect. This leads to relatively higher \Zgt{} in the centres of galaxies and consequently steeper \Zgt{} gradients by $z=0$.

We note here that AGN feedback has a relatively minor impact on the metallicity profiles in disc-dominant systems in \lgaltt{}. This is because it is most effective in galaxies with large supermassive black holes (SMBHs) and hot gas reservoirs, acting to reduce or prevent further gas accretion onto the disc (see section S1.14 of the Supplementary Material). For star-forming main-sequence galaxies, such as those in our sample here, SMBH masses and their accretion rates are relatively low, leading to a relatively mild suppression of their net cooling rates (which remain on average at $\sim{}1-10\ \Msun/\tn{yr}$ by $z=0$, see also \citealt{Yates+14}). However, alternative forms of AGN feedback formalism, which allow significant amounts of gas to be ejected from star-forming galaxies via AGN-driven outflows, could affect the metallicities in the ISM and CGM, and therefore also the evolution of disc metallicity profiles.

There have also been a number of observational studies of metallicity profiles at higher redshift recently (\eg{}\citealt{Troncoso+14,Stott+14,Leethochawalit+16,Curti+20b,Wang+20,Gillman+21}). In general, these studies suggest that (a) profiles are rather flat at all redshifts back to $z\sim{}2.5$, albeit with a wide spread in $\Delta(\Zg{})$ (see \citealt{Hemler+20}), and (b) strong, metal-rich galactic outflows are required in order to explain this. These findings are in reasonable qualitative agreement with the modified \lgaltt{} model, which allows highly efficient metal removal from galaxies via outflows. For example, when selecting galaxies \textit{at fixed mass} (rather than as direct progenitors of present-day galaxies), and measuring $\Delta(\Zg{})$ in dex/kpc across a radial range of $\sim{}0-10$ kpc (as is typically the case for higher-redshift observations), we find that for galaxies with $9.6 \leq \logMm{} < 10.2$ that $\Delta(\Zg{}) \sim{}-0.08$ dex/kpc on average at $z\sim{}2$, flattening only mildly to $-0.04$ dex/kpc by $z=0$.

However, the heterogeneity and limited resolution of the currently available observational data at high redshift make more detailed comparisons with models difficult. The measured \Zgt{} slope in an observed high-redshift galaxy depends sensitively on a number of often uncontrollable or uncertain factors, such as its inclination, morphology, the radial range covered, radial resolution, normalisation chosen (\eg{}$R$/\Reff{} or $R$/kpc), and complex sample selection biases. Additionally, some SL \Zgt{} diagnostics are expected to be less applicable to high-redshift galaxies than others, due to differences between the gas properties within these high-redshift systems and those at low redshift which are used for calibration (\eg{}\citealt{Cullen+14,Strom+18}). We therefore choose not to force a more detailed comparison between \Zgt{} profiles at high redshift in observations and models in this work. The arrival of instruments such as JWST/NIRSpec, VLT/ERIS, and VLT/MOONS will facilitate a much more precise, accurate, and comprehensive comparison between models and observations in future.

\subsection{Alternative model variants}\label{sec:Alternative model variants}
\subsubsection{A gas-density-dependent CGM enrichment efficiency}\label{sec:SigGas-dep model}
One way to further refine the outflow prescription in \lgaltt{} is to tie the CGM enrichment efficiency to the density of the ISM gas through which outflows must pass. Such approaches have been trialled before in the \lgal{} and \textsc{Galform} semi-analytic models \citep{Yates+13,Lagos+13}. Here, we explore an extension to this formalism, by setting the amount of direct CGM enrichment from SNe to be inversely proportional to the log of the local ISM density in each radial ring. The effective CGM enrichment efficiency of both SNe-II and SNe-Ia is then given by,
\begin{equation}\label{eqn:GasDen-dep fwind}
    f\sub{SNe,hot} = \tn{min}\left[1.0\,,\,\frac{1.5}{\logten{}(\Sigma\sub{ISM}/\Msun{}\,\tn{pc}^{-2})}\right]\;\;, 
\end{equation}
where the value of the normalisation constant, $10^{1.5}\approx{}32\Msun{}\,\tn{pc}^{-2}$, is chosen to reflect the typical gas density below which the H$_{2}$ fraction drops to zero according the \citet{Krumholz+09} prescription used in \lgaltt{} (see Section \ref{sec:L-Galaxies}). Above this density, $f\sub{SNe,hot}$ decreases, reaching $\sim{}50$ per cent at ISM densities of $\sim{}1000\Msun{}\,\tn{pc}^{-2}$.

\begin{figure*}
\centering
 \includegraphics[angle=0,width=0.99\linewidth]{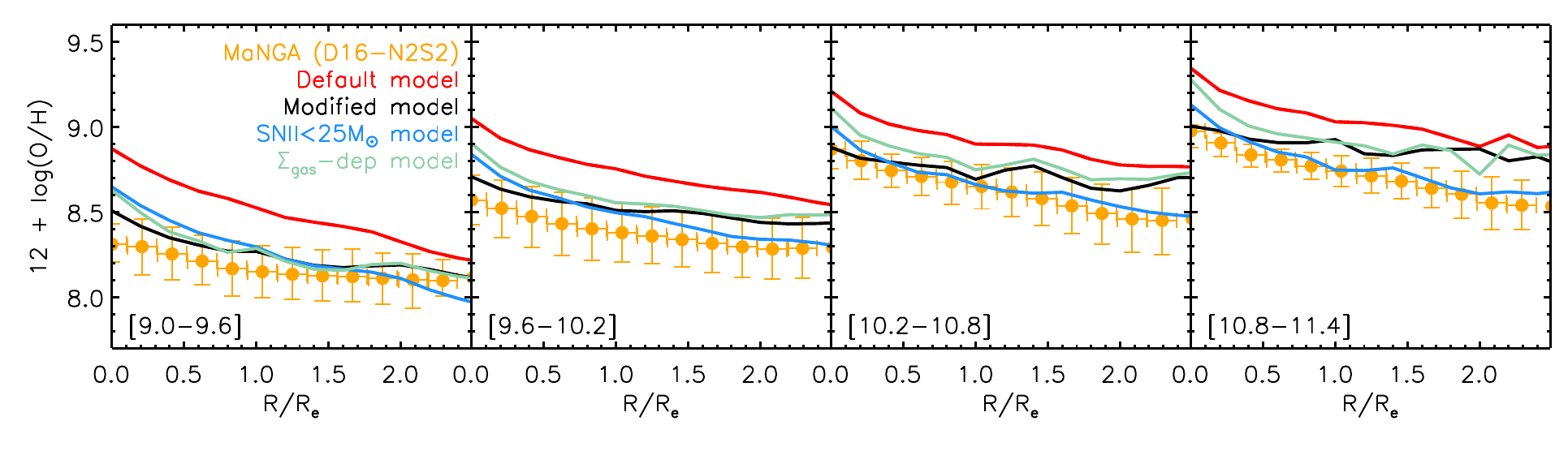}
 \caption{Mean radial \Zgt{} profiles out to 2.5\Reff{} for galaxies in four mass bins. The default (red), modified (black), $\Sigma\sub{ISM}$-dependent (green, see Section \ref{sec:SigGas-dep model}), and MAXGCEMass25 (blue, see Section \ref{sec:Low-SNII-max model}) models are shown, along with our MaNGA sample (orange circles).}
 \label{fig:AllModels_Zg_profiles_z0}
\end{figure*}



The mean \Zgt{} profiles for this `$\Sigma\sub{ISM}$-dependent model' (green) are shown in Fig. \ref{fig:AllModels_Zg_profiles_z0}. This version of \lgaltt{} performs similarly to the modified model presented earlier in this work, except that it returns steeper slopes in the centres of galaxies due to the higher-density ISM present in these regions.\footnote{In \lgaltt{}, ISM densities exceeding $32\Msun{}\,\tn{pc}^{-2}$ are typically found within $\sim{}1.2$ R$\sub{e}$ for disc galaxies at $z=0$.} This brings the $\Sigma\sub{ISM}$-dependent model into worse agreement with observations in terms of \Zgt{} profile normalisation, but somewhat better agreement in terms of the median slope (see Section \ref{sec:Mass dependent Zg profiles}).

The $\Sigma\sub{ISM}$-dependent model fails to adequately reproduce the evolution of global metallicity at fixed mass. As in the default model, a slightly \textit{inverse} evolution occurs at low mass due to the over-enrichment of the ISM at very early times followed by metal-poor accretion and radial inflow thereafter. We therefore conclude that highly-efficient direct CGM enrichment is also favoured in the denser centres of galaxies. This result could be compatible with the findings of some high-resolution ISM simulations (\eg{}\citealt{Gatto+17}), in which pre-SN stellar winds and radiation are able to efficiently clear dense ambient gas away from SN sites, allowing for efficient SN-driven galactic winds to be driven (see Section \ref{sec:Comparison to other models}).

\subsubsection{A reduced maximum SN-II progenitor mass}\label{sec:Low-SNII-max model}
Another way to reduce the \Zgt{} in the ISM of model galaxies is to decrease the overall amount of oxygen produced in the universe. This can be most directly done by lowering the upper mass limit for stars that can contribute to chemical enrichment. In \lgaltt{}, massive stars are assumed to contribute via stellar winds and SNe-II up to a maximum birth mass of $M\sub{GCE,max} = 120\Msun$. This limit was chosen due to the fact that such massive stars are known to exist in the real Universe (see \citealt{Cote+16}, and references therein), and that they are required by some chemical evolution models to match the element ratios seen in stars and \HII{} regions in the solar neighbourhood (\eg{}\citealt{Portinari+98,Francois+04,Carigi+20}). However, other observational studies have suggested that the upper mass limit for SN-II progenitors should be considerably lower (\eg{}\citealt{Smartt+15,Davies&Beasor18,Schady+19}), and other chemical evolution models have been successful in reproducing Milky Way constraints without requiring enrichment from stars above $\sim{}25-50\Msun$ (\eg{}\citealt{Kobayashi&Nakasato11,Cote+17}).

Therefore, we assess the impact of lowering the mass limit of SN-II progenitors in \lgaltt{}, by running a variant of the model with $M\sub{GCE,max}=25\Msun$. This value was chosen to roughly mimic the findings of \citet{Ertl+16}, which suggest that stars of mass $8\lesssim \tn{\logM{}} \lesssim{}22$ are relatively efficient at forming SNe-II, whereas at higher masses there are only `islands of explodability', with most stars directly collapsing into black holes. In this `MaxGCEMass25' model, we maintain the upper mass threshold of the IMF at $120\Msun$, implying that stars between 25 and $120\Msun$ are formed but do not contribute at all to chemical enrichment or SN-driven winds. We also fix the values of all other \lgaltt{} parameters to those used in the default model, including the GCE parameters listed in Table \ref{tab:GCE parameters}.

Fig. \ref{fig:AllModels_Zg_profiles_z0} shows the radial \Zgt{} profiles for this MaxGCEMass25 model as blue lines. We find a remarkable agreement in normalisation, slope, and scatter between the MaxGCEMass25 model and MaNGA data at high stellar mass. Although, we note that a shallower slope in high-mass galaxies is favoured by higher-resolution MUSE data (see Section \ref{sec:Profiles at z=0}). The \Zst{} profiles (not shown here) also reveal a good correspondence, with the MaxGCEMass25 model exhibiting slightly steeper slopes than the modified model out to large radii, in all mass bins. This is due to the lack of highly-enriched accretion onto the outskirts of galaxy discs in the MaxGCEMass25 model, as the $f\sub{hot}$ parameter values are low.


A key difference between the MaxGCEMass25 and modified models should be the chemical abundances found \textit{outside} galaxies, as these should differ greatly depending on whether massive stars are allowed to efficiently enrich the CGM or not at all. In Fig. \ref{fig:TZRs_ICM_incSNIIMAX25}, the T\ZICMt{}R for the MaxGCEMass25 model (blue lines) is shown alongside those of the default and modified models. [Mg/H] is clearly under-predicted by $\sim{}0.5$ dex in the MaxGCEMass25 model relative to the other \lgaltt{} model variants and observations, even when considering the issues with Mg measurements discussed in Section \ref{sec:Zh}. This under-abundance is caused by both the production and ejection of light $\alpha$ elements being greatly reduced in the MaxGCEMass25 model compared to the modified model. Heavier $\alpha$ elements, such as Si and S, are less affected by changes to $M\sub{GCE,max}$, because they are predominantly produced in lower-mass SN-II progenitor stars (see \citealt{Portinari+98}) which are allowed to enrich in all the versions of \lgaltt{} considered here.

Like the $\Sigma\sub{ISM}$-dependent model discussed above, the MaxGCEMass25 model also exhibits a lack of evolution in the M\Zgt{}R and M\Zst{}R over cosmic time, due to an over-retention of oxygen inside galaxies. This is in contradiction with the observations discussed in Section \ref{sec:Z_evo}.

These shortcomings can be improved somewhat by increasing the value of $f\sub{SNII,hot}$ in the MaxGCEMass25 model, in line with what is already done in the modified model. However, we find that even when doubling the value of $f\sub{SNII,hot}$ to 0.6, the [Mg/H] in the ICM is still too low by $\sim{}0.4$ dex, and there is still effectively no evolution in \Zgt{} over time at low mass. This suggests that there simply isn't enough metal produced in the MaxGCEMass25 model considered here to adequately reproduce the chemical content observed in and around galaxies.

Allowing stars above $25\Msun$ to shed metal-rich winds \textit{before} directly collapsing into black holes could help increase the overall metal budget in the MaxGCEMass25 model (see also \citealt{Kobayashi+20,Gutcke+20}). The combination of such pre-collapse wind enrichment with higher values of $f\sub{SNII,hot}$ will be an avenue of investigation with \lgaltt{} in future work.



\section{Comparisons to other models}\label{sec:Comparison to other models}
\subsection{Metal-rich winds}\label{sec:Comps:Metal-rich winds}
The main finding in this work, that a large fraction of the ejecta from SNe needs to be deposited directly into the CGM surrounding star-forming galaxies, is in good correspondence with earlier \lgal{} work by \citet{Fu+13} as well as a number of other theoretical studies. For example, \citet{Gibson+13} have found that metal-rich gas is efficiently removed from the centres of their two model $L^{*}$ galaxies (the MaGICC simulations) when implementing an enhanced feedback prescription which enables 100 per cent of available SN energy to couple with the surrounding gas. This leads to flattened metallicity profiles at early times in these systems, similar to a sub-set of the high-redshift observations to which they compare. Metallicity profiles are similarly flattened in \lgaltt{} when switching from the default model to the modified model, with the latter being more efficient at removing metals from galaxies.

Similarly motivated by observed high-redshift \Zgt{} slope estimates, \citet{Ma+17} show that the efficient SN feedback implemented into the \textsc{Fire} simulations \citep{Hopkins+14} causes flattened metallicity profiles. Of their five model galaxies with $\logMm{}\gtrsim{}10.0$, four exhibit weakly-negative \Zgt{} slopes across a similar radial range to that considered in this work, with $\Delta{}(\Zg{}) \lesssim -0.07$ dex/kpc. This is in reasonable agreement with high-mass galaxies in \lgaltt{}.

Despite the similarity between the results of these various models, there is an interesting difference in their feedback prescriptions. In MaGICC and \textsc{Fire}, feedback is enhanced by allowing more material from the ISM to be entrained in galactic outflows, leading to increased CGM enrichment \textit{and} increased mass-loading factors (see discussion by \citealt{Mitchell+20a}). Conversely, in the modified \lgaltt{} model, enhanced CGM enrichment occurs \textit{without} a corresponding increase in the mass-loading factor. This can be seen in Figs. \ref{fig:Reheating_and_ejection_rates} and \ref{fig:mass_loading_factors}, which show that the increased metal deposition into the CGM is accompanied by slightly lowered overall reheating rates and mass-loading factors, due to the reduced coupling between SNe ejecta and the ambient ISM. Physically speaking, this could be interpreted as low-density chimneys being opened-up in star-forming galaxies by on-going SN feedback.

Such a picture is seen in the high-resolution, stratified-disc simulations from the SILCC project \citep{Walch+15}, which have a spatial cell resolution of $<10$ pc and initial gas surface density of $\Sigma\sub{gas} = 10\,\Msun/\tn{pc}^{2}$. These  simulations show that pre-SN stellar winds and ionising radiation can lower the density around the sites of SNe progenitors, allowing efficient galactic winds to be driven when the volume filling factor of hot gas in the ISM exceeds $\sim{}50$ per cent \citep{Gatto+17,Peters+17}. Mass-loading factors typically reach 1-10 in SILCC \citep{Girichidis+16,Gatto+17}, in good agreement with those found in \lgaltt{} (see Fig. \ref{fig:mass_loading_factors}). In their simulations of similar size and spatial resolution to SILCC, \citealt{Li+17} also find that 40-90 per cent of the metals ejected by SNe end up leaving the galaxy directly in such outflows.


%

These findings all point to a conclusion that \textit{high-mass} galactic outflows are not necessarily required to reproduce the metallicities seen in main-sequence galaxies. Instead, more moderate mass-loading factors, accompanied by efficient \textit{direct} CGM enrichment by SNe, can provide an equally viable solution.

\subsection{MZR evolution}\label{sec:Comps:MZgR evolution}
\citet{Collacchioni+18} have found that an explicit redshift dependence in the reheating rate is required to drive an evolution in \Zgt{} at fixed mass in the \textsc{Sag} semi-analytic model. This modification boosts SN feedback in order to increase the amount of ISM gas reheated at high redshift. The original, unmodified prescriptions for calculating reheating and ejection energies in \textsc{Sag} are similar to those used in \lgal{}. Both formalisms rely on a dependence on the DM subhalo velocity, which naturally provides an evolution with both mass and redshift (see Fig. \ref{fig:Reheating_and_ejection_rates}). Nonetheless, this alone is not enough to drive an evolution in the M\Zgt{}R in our default model, in agreement with the findings of \citet{Collacchioni+18}. However, such a reheating formalism \textit{is} sufficient when coupled with a high, fixed direct CGM enrichment efficiency in our modified model. Our approach also has only a minimal effect on the cosmic SFRD evolution, in contrast to the modified \textsc{Sag} set-up (see their fig. 6).

Similarly to \textsc{Sag}, \citet{Lian+18a,Lian+18c} are only able to simultaneously reproduce an evolution in the M\Zgt{}R and M\Zst{}R for their stand-alone GCE model when implementing a strongly time- and mass-dependent $f\sub{hot}$ parameter. We note that part of the motivation for such an approach is the large difference between the ISM and stellar metallicities inferred by \citet{Lian+18a} from their chosen low-redshift observations. This difference, reaching $\sim{} 0.8$ dex at low mass, requires particularly strong direct metal ejection at early times, with a value of $f\sub{hot} = 1.0$ being maintained over the first 9 Gyr for galaxies with \logM{} $= 9.0$. Due to the smaller difference between \Zgt{} and \Zst{} of $\sim{}0.2$ dex we find for our MaNGA sample when using the D16-N2S2 diagnostic to calculate \Zgt{}, as well as the additional astrophysics modelled in \lgaltt{}, we find that such explicit time and mass dependencies of $f\sub{hot}$ are not required in our models.

A mass (although not time) dependent $f\sub{SNII,hot}$ parameter was found by \citet{Somerville+15} to drive an M\Zgt{}R evolution in the \textsc{Santa Cruz} semi-analytic model, when studying various combinations of H$_{2}$ formation, star formation, and metal outflow recipes. A model variant in which $f\sub{SNII,hot} \propto{} \tn{exp}(-M\sub{vir})$ was found to reduce star formation at early times, and therefore drive a greater evolution in \Zgt{} below $z\sim{}4$. Many of the other combinations considered returned a negligible or even inverted evolution in \Zgt{} with cosmic time, as is also seen for our default model.

Finally, an evolution in the M\Zgt{}R is seen in the \textsc{Illustris-TNG100} hydrodynamical model despite having an \textit{increase} in the metal reheating efficiency over cosmic time, especially in higher-mass galaxies \citep{Torrey+19}. That study determined that \Zgt{} evolution is primarily driven by changes in the ISM gas fraction, rather than the efficiency of metal removal via SN feedback.

The range of different approaches taken by the above models to reproduce the M\Zgt{}R suggests that additional constraints, such as the evolution of metallicity profiles, are also important when trying to pin-down the key physical processes actually driving galactic chemical evolution.

\subsection{Metallicity profiles}\label{sec:Comps:Metallicity gradients}
Recently, \citet{Tissera+19} found relatively flat slopes at $z=0$ for their 592 model galaxies from the EAGLE hydrodynamical simulation. They report a weak correlation between $\Delta{}(\Zg{})$ and stellar mass for secularly-evolving galaxies when measuring $\Delta{}(\Zg{})$ in dex/kpc, and identify inside-out growth as one of the mechanisms driving flatter \Zgt{} profiles in massive galaxies. Overall, these EAGLE results are in qualitative agreement with the findings from \lgaltt{} presented here, reflecting the similar chemical enrichment schemes implemented into the two models.

When extending their GCE model to account for metallicity gradients, \citet{Lian+18b,Lian+19} found that an additional radial dependence on the $f\sub{hot}$ parameter enabled a better match to the slopes and normalisations of their preferred observed \Zgt{} and \Zst{} profiles at $z=0$. The main differences between the MaNGA data set used by \citet{Lian+18b,Lian+19} and that considered here are (a) the choice of SL \Zgt{} diagnostic, (b) the use of VOR10 versus HYB10 datacubes for the gas-phase analysis, and (c) our spiral morphology requirement. When comparing to our preferred observational sample, the modified \lgaltt{} model does not require an additional radial dependence on $f\sub{hot}$.

Finally, \citet{Hemler+20} have recently investigated the evolution of \Zgt{} profiles in the \textsc{Illustris-TNG50} simulation. At low redshift, there is good qualitative agreement between their findings and those of the modified \lgaltt{} model. For example, \citet{Hemler+20} find a weak correlation between stellar mass and $\Delta{}(\Zg{})$ (when measured in dex/kpc) at $z=0$. However, they report a consistent flattening of average \Zgt{} profiles over cosmic time \textit{at fixed mass} in \textsc{Illustris-TNG50}. In contrast, \Zgt{} profile slopes remain fairly constant over cosmic time at fixed mass in \lgaltt{}, as discussed in Section \ref{sec:Profiles evo}. For example, the median $\Delta{}(\Zg{})$ changes from -0.042 dex/kpc at $z=2$ to -0.037 dex/kpc at $z=0$ for the mass bin $9.6 \leq{} \logMm{} < 10.2$ in our modified model, when measured between 0.5 and 2\Reff{}. This puts \lgaltt{} in better agreement with models such as MaGICC and \textsc{Fire}. When instead tracing the change in $\Delta{}(\Zg{})$ for \textit{the same galaxies} as they evolve over cosmic time, we do find a consistent flattening in \lgaltt{}, as shown in Fig. \ref{fig:Zg_profiles_perRgc_evo}. This highlights the importance of distinguishing between fixed-mass populations and individual galaxies when assessing and comparing the evolution of \Zgt{} profiles.

\section{Conclusions}\label{sec:Conclusions}
In this work, we present a modified version of the \lgaltt{} galaxy evolution model. This modified model performs better than our default model in reproducing the gas and stellar radial metallicity profiles observed in low-redshift star-forming disc galaxies, as well as the present-day galaxy mass -- metallicity relations for gas and stars, and their evolution back to $z\sim{}2-3$.These improvements are achieved without significantly altering the good correspondence that \lgaltt{} has with other key galaxy population properties such as the stellar mass function, H\textsc{i} mass function, $M_{*}$ -- sSFR relation, and cosmic SFR density evolution. In order to compare with IFU observations at low redshift, we have formed and derived radial profiles for a sample of 571 MaNGA star-forming disc galaxies, and also utilised MUSE galaxy profiles made available in the literature. The following key results are obtained:

\begin{itemize}
    \item Highly-efficient direct enrichment of the CGM by SNe-II is required in \lgaltt{} to reproduce \Zgt{} and \Zst{} profiles in nearby disc galaxies. $\sim{}90$ per cent of SN-II ejecta directly enriching the CGM is preferred, with $\sim{}80$ per cent for SNe-Ia (see Section \ref{sec:Profiles at z=0}). We find that values below $\sim{}75$ per cent lead to an overly-enriched ISM and stellar disc in galaxies at both high and low redshift.
    
    \item Such high CGM enrichment efficiencies also enable \lgaltt{} to reproduce the stellar and gas-phase mass -- metallicity relations at $z=0$ and their evolution back to $z\sim{}2-3$, as well as the metal content found in the hot gas surrounding star-forming galaxies, groups, and clusters (see Sections \ref{sec:MZRs at z0} \& \ref{sec:Z_evo}).
    
    \item We find a weak correlation between stellar mass and \Zgt{} profile slope in the modified model at $z=0$, such that more massive galaxies tend to have flatter profiles. This is in general agreement with both our MaNGA sample above \logM{} $\sim{}10$ and other recent IFU studies (see Section \ref{sec:Mass dependent Zg profiles}).
    
    \item We note that this trend between stellar mass and \Zgt{} profile slope in observational samples is sensitive to both the metallicity diagnostic and galaxy morphology cut chosen, particularly for low-mass systems (see Section \ref{sec:Mass dependent Zg profiles}). Slope measurements are also sensitive to the particular radial range and radial scaling chosen.
    
    
    
    \item The weakly mass-dependent \Zgt{} slopes seen in \lgaltt{} are primarily caused by more efficient metal enrichment at large radii in massive galaxies than in low-mass galaxies. This enrichment is driven by a combination of inside-out growth and metal-rich accretion (see Section \ref{sec:Profiles evo}).
    
    
        
    
    \item We find that gas-density-dependent direct CGM enrichment efficiencies are also compatible with observed \Zgt{} and \Zst{} profiles. However, values above $\sim{}60$ per cent are still required in \lgaltt{}, even for high-density regions near the centres of galaxies (see Section \ref{sec:SigGas-dep model}).
    
    \item Reducing the upper mass limit for SN-II progenitor stars to $25\Msun$ also reproduces the \Zgt{} and \Zst{} profiles seen in low-redshift star-forming galaxies, without requiring strong direct CGM enrichment. However, this model variant severely under-predicts the magnesium abundances found in the intragroup and intracluster medium, as well as the observed evolution of the M\Zgt{}R (see Section \ref{sec:Low-SNII-max model}).
\end{itemize}

In future work, we intend to study the effects of combining high efficiencies of metal ejection into the CGM with a reduced maximum SN-II progenitor mass in \lgaltt{}, in tandem with a re-evaluation of the input stellar yield tables used. This will allow us to further constrain the chemical properties of both star-forming and quiescent galaxies at high redshift.
\section*{Data availability}
The \lgaltt{} source code, as well as example output catalogues from the default and modified models, are publically available at \href{https://lgalaxiespublicrelease.github.io/}{lgalaxiespublicrelease.github.io/}. Complete output catalogues for the \lgaltt{} default model are also available on the \textsc{Millennium} database at \href{http://gavo.mpa-garching.mpg.de/Millennium/}{gavo.mpa-garching.mpg.de/Millennium/}. Additional \lgal{} data and the MaNGA derived data products presented here can be obtained from the main author upon request.

\section*{Acknowledgements}
The authors would like to thank the referee for very helpful and constructive comments and suggestions, as well as Michael Blanton, Payel Das, Thales Gutcke, Fran\c{c}ois Mernier, C\'{e}line P\'{e}roux, Doris Stoppacher, David Wake, Jessica Werk, Jabran Zahid, and Almudena Zurita, for valuable discussions during the undertaking of this work. BMBH acknowledges support from a Zwicky Prize fellowship. JF acknowledges the support by the Youth innovation Promotion Association CAS and Shanghai Committee of Science and Technology grant No.19ZR1466700. Part of this research was carried out on the HPC resources at the MPCDF in Garching operated by the Max Planck Society (MPG).



\bibliographystyle{mnras}
\bibliography{robyates.bib}







\bsp	
\label{lastpage}
\end{document}